\documentclass[iop,apj,numberedappendix]{emulateapj}
\slugcomment{{\sc Accepted to ApJ}}
\usepackage{natbib}
\usepackage{graphicx}
\usepackage{amsmath,amsfonts}
\usepackage{multirow}
\usepackage{mathtools}
\usepackage{hyperref}
\usepackage{placeins}
\usepackage{afterpage}

\setcitestyle{notesep={; }}
\def\sgra{Sgr~A$^{\ast}$}

\defcitealias{TMS}{TMS}

\begin{document}
\title{Interferometric Imaging Directly with Closure Phases and Closure Amplitudes}
\shorttitle{Closure-Only Interferometric Imaging}

\author{
Andrew A. Chael\altaffilmark{1},
Michael D. Johnson\altaffilmark{1},
Katherine L. Bouman\altaffilmark{1,2},
Lindy L. Blackburn\altaffilmark{1},
Kazunori Akiyama\altaffilmark{3,4,5},
Ramesh Narayan\altaffilmark{1}
}
\shortauthors{Chael et al.}
\altaffiltext{1}{Harvard-Smithsonian Center for Astrophysics, 60 Garden Street, Cambridge, MA 02138, USA}
\altaffiltext{2}{Massachusetts Institute of Technology, Computer Science and Artificial Intelligence Laboratory, 32 Vassar Street, Cambridge, MA 02139, USA}
\altaffiltext{3}{National Radio Astronomy Observatory, 520 Edgemont Rd, Charlottesville, VA, 22903, USA}
\altaffiltext{4}{Massachusetts Institute of Technology, Haystack Observatory, Route 40, Westford, MA 01886, USA}
\altaffiltext{5}{National Astronomical Observatory of Japan, 2-21-1 Osawa, Mitaka, Tokyo 181-8588, Japan}
\email{achael@cfa.harvard.edu}

\begin{abstract}
Interferometric imaging now achieves angular resolutions as fine as ${\sim}10~\mu$as, probing scales that are inaccessible to single telescopes. Traditional synthesis imaging methods require calibrated visibilities; however, interferometric calibration is challenging, especially at high frequencies. Nevertheless, most studies present only a single image of their data after a process of ``self-calibration,'' an iterative procedure where the initial image and calibration assumptions can significantly influence the final image. We present a method for efficient interferometric imaging directly using only closure amplitudes and closure phases, which are immune to station-based calibration errors. Closure-only imaging provides results that are as non-committal as possible and allows for reconstructing an image independently from separate amplitude and phase self-calibration. While closure-only imaging eliminates some image information (e.g., the total image flux density and the image centroid), this information can be recovered through a small number of additional constraints. We demonstrate that closure-only imaging can produce high fidelity results, even for sparse arrays such as the Event Horizon Telescope, and that the resulting  images are independent of the level of systematic amplitude error. We apply closure imaging to VLBA and ALMA data and show that it is capable of matching or exceeding the performance of traditional self-calibration and CLEAN for these data sets. 
\end{abstract}


\section{Introduction}
Synthesis imaging for interferometry is an ill-posed problem. An interferometer measures a set of complex visibilities which sample the Fourier components of an image. Standard deconvolution approaches to imaging, such as the CLEAN algorithm \citep{Hog_1974}, begin with an inverse Fourier transform of the sampled visibilities and then proceed to deconvolve artifacts introduced by the sparse sampling in the Fourier domain. To use CLEAN and other traditional imaging algorithms, interferometric visibilities must be calibrated for amplitude and phase errors. However, at high frequencies, the atmospheric coherence time can be as short as seconds, introducing rapid phase variations and effectively eliminating the capability to measure absolute interferometric phase. Amplitude calibration also becomes more difficult at high frequencies, and pointing errors due to small antenna beam sizes can introduce large, time-varying errors in visibility amplitudes. While amplitude gain errors typically have longer characteristic timescales than phase errors, some Very Long Baseline Interferometry (VLBI) instruments such as the Event Horizon Telescope \citep[EHT;][]{Decadal} use phased arrays as single stations, which can introduce rapid amplitude variations from fluctuations in the individual station's phasing efficiency. 

A key simplification for interferometric calibration arises due to the fact most calibration errors can be decoupled into station-based gain errors \citep[e.g.,][]{Hamaker_1996,TMS}. For an interferometric array consisting of $N$ sites, there are $N(N-1)/2$ visibilities at each time, but only $N$ unknown gains. Hence, the calibration is over-constrained and combinations of visibilities can be formed that are unaffected by calibration errors. For example, a closure phase is the phase of a product of three visibilities around a triangle, which cancels out the station-based phase errors on each individual visibility \citep{Jennison_1958,Rogers_74}. Likewise, the closure amplitude is a combination of four visibility amplitudes that cancels out amplitude gain errors in a specified ratio \citep{Twiss_60, TMS}. Both of these quantities provide access to information about the source image that is unaffected by calibration assumptions. Despite the challenges in \emph{absolute} calibration of a VLBI array, closure quantities provide robust measurements of certain \emph{relative} quantities, which carry information about source structure that is only limited by the level of thermal noise. 

The standard algorithm used for interferometric imaging is CLEAN \citep{Hog_1974, Clark_80}, which deconvolves a dirty image produced by an inverse Fourier transform by decomposing it into point sources. When the calibration is uncertain, the usual approach is to iterate between imaging with CLEAN and deriving new calibration solutions using information from the previous image --- a so-called ``self calibration'' or ``hybrid mapping'' loop \citep[e.g.,][]{Wilkinson77,Readhead78,Readhead80,Schwab_1980, Cornwell_1981,Pearson_1984,Walker_1995,Cornwell_1999,TMS}. 
The results and time to convergence of this approach depend on many assumptions made in the course of this hybrid process, including the initial source model used for self-calibration, the choice of which regions to clean in a given iteration,  the method used for deriving complex gains from a given image, and the choice of how frequently to re-calibrate the data. The sensitivity of the final image to these assumptions cannot be directly inferred from the result. 

In contrast with CLEAN's approach of deconvolving the dirty image into point sources, another family of methods (most famously the Maximum Entropy Method, or MEM, see e.g \citep{NN_1986}) for interferometric imaging solves directly the source image pixels by fitting them to data, constrained by additional convex regularization terms, such as entropy, sparsity, or smoothness \citep[e.g.,][]{Frieden_1972,GS_1978,Cornwell_1985,Briggs_NNLS}. Like CLEAN, these MEM-like methods can be used to produce images from complex visibilities in conjunction with a self-calibration loop. In contrast with CLEAN, however, these approaches can also be used directly with other data products derived from complex visibilities, as they only rely on comparing the data computed from the reconstructed image to the specified measurements. In  other words, these methods never need to perform an inverse Fourier transform from calibrated input data.  Consequently, these approaches can use closure quantities directly as the fundamental data product, bypassing the self-calibration loop entirely. The field of optical interferometry, for example, has pioneered the use of imaging directly from the measured visibility amplitudes and closure phases, bypassing the corrupted visibility phase \citep{BSMEM_94, SQUEEZE, Thiebaut_2013, Thiebaut_2017}. Recently, several other methods have built on these techniques in preparing imaging algorithms for EHT data, fitting some combination of closure phases and visibility amplitudes directly while using different regularizing functions \citep{Katie_2015,akiyama_bs}.

In this paper, we take the next step and present a method to reconstruct images directly using only closure amplitudes and closure phases. Our reconstructions require no assumptions about absolute phase or amplitude calibration beyond stability during the integration time used to obtain the visibilities. Though we find that a single round of self-calibration to the final image and re-imaging with complex visibilities can produce even better results,\S\ref{sec::ALMA}. To make closure-only imaging computationally efficient, we derive analytic gradients of the data chi-squared terms for closure quantities, which greatly improves the speed of our algorithm. When using these analytic gradients, closure-only imaging of VLBI data does not require significantly more computational time than standard imaging with complex visibilities, and it is still feasible on a personal computer for large datasets (e.g., those of connected-element interferometers such as ALMA). 

We begin, in \S\ref{sec::c_quantities}, by reviewing fundamental properties of interferometric visibilities and closure quantities. Next, in \S\ref{sec::Imaging}, we discuss imaging via regularized maximum likelihood and demonstrate how to efficiently implement closure-only imaging in this framework. In \S\ref{sec::imp}, we detail our implementation of closure-only imaging in the \texttt{eht-imaging} software library\citep{ehtim},\footnote{Codebase: \url{https://github.com/achael/eht-imaging}} our methods for simulating data with gain and phase errors, and our techniques for evaluating the fidelity of the reconstructed images. In \S\ref{sec::Results}, we show the results of applying our method to both simulated EHT data and real datasets from the VLBA and ALMA. In \S\ref{sec::Discussion}, we discuss general properties of closure-only imaging, and in \S\ref{sec::Conclusions}, we summarize our results.

\section{Visibilities and Closure Quantities}
\label{sec::c_quantities}

\subsection{Interferometric Visibilities}
The van Cittert-Zernike theorem identifies the visibility $V_{ij}$ measured by an baseline $\vec{b}_{ij}$ between stations $i$ and $j$ as a Fourier component of the source image intensity distribution $I(x,y)$  \citep[][hereafter TMS]{TMS}:
\begin{align}
 \label{eq::VCZ}
 V_{ij}= \tilde{I}(u,v) = \int\int\, I(x,y)e^{-2\pi \mathrm{i} (ux+vy)}\, \mathrm{d}x\,\mathrm{d}y.
\end{align}
Here, $x$ and $y$ are real space angular coordinates and $u$, $v$ are the coordinates of the given baseline vector $\vec{b}_{ij}$ projected in the plane perpendicular to the line of sight and measured in wavelengths. Since $I(x,y)$ is a real number, the visibility is conjugate-symmetric in the Fourier plane, $\tilde{I}(-u,-v) = \tilde{I}^*(u,v)$. When $N_s$ stations can observe the source, the number of independent instantaneous visibilities is given by the binomial coefficient
\begin{equation}
\label{eq::nvis}
 N_\text{vis} = {N_s\choose2} = \frac{N_s(N_s-1)}{2}.
\end{equation}
To fill in samples of the Fourier plane from the small number $N_{\rm vis}$ available at a single instant in time, interferometric observations typically use the technique of ``earth rotation aperture synthesis.'' As the Earth rotates, the projected baseline coordinates $(u,v)$ trace out elliptical curves in the Fourier domain, providing measurements of new visibilities.

The identification of measured visibilities with Fourier components of the image is complicated by several factors. First, thermal noise from the telescope receiver chains, Earth's atmosphere, and astronomical background is added to the measured visibility. This thermal noise, $\epsilon_{ij}$, is Gaussian with a time- and baseline-dependent standard deviation, which depends on the telescope sensitivities, bandwidth, and integration time. Second, each station transforms the measured incoming polarized waveform according to its own (time-dependent) $2{\times}2$ Jones matrix that adjusts the level of the measured signal amplitude and mixes the measured polarizations (e.g.\citet{Hamaker_1996}\citetalias{TMS}).  For the purposes of this paper, we ignore polarization and consider each station as contributing a single (time-dependent) complex gain $G_ie^{\mathrm{i}\phi_i}$ to the visibility. 

The phase error $\phi_i$ results from uncorrected propagation delays and clock errors. In particular, atmospheric turbulence contributes a rapidly varying stochastic term to each $\phi_i$, which generally varies more quickly than the amplitude gain term, $G_i$, which arises from uncertainty in the conversion of the correlation coefficients measured on each baseline to units of flux density. In effect, the visibility amplitude is first measured in units of the noise, and then scaled to physical units from knowledge of the telescope noise properties.

Including all of these corrupting factors, the full complex visibility is
\begin{equation}
\label{eq::gains}
 V_{ij} = G_iG_je^{\mathrm{i}(\phi_i-\phi_j)}\left(\tilde{I}(u,v) + \epsilon_{ij}\right),
\end{equation}
where the gain amplitudes, phases, and thermal noise all vary in time. 

Note that Equation~\eqref{eq::gains} represents all systematic errors (e.g., those other than thermal noise) as station-based effects. In practice, effects such as polarization leakage and bandpass errors will also contribute small baseline-based effects that can bias closure quantities. However, these errors are generally slowly varying and can be removed with \emph{a priori} calibration.

\subsection{Closure Phases and Closure Amplitudes}
Two types of ``closure quantities'' can be formed from complex visibilities that are insensitive to the particular station-based complex gain terms. While these quantities are robust to the presence of arbitrarily large complex gains on the visibilities, they contain less information about the source than the full set of complex visibilities. Furthermore, because closure quantities mix different Fourier components they can be difficult to interpret physically.

First, multiplying three complex visibilities around a triangle of baselines eliminates the complex gain phase terms (\citet{Jennison_1958,Rogers_74},\citetalias{TMS}). For any 3 stations, the visibility \emph{bispectrum} is  
\begin{equation}
 \label{eq::bispec}
 V_B \equiv |V_B|e^{\mathrm{i}\psi} = V_{12}\,V_{23}\,V_{31}.
\end{equation}
While the bispectral amplitude $|V_B|$ is affected by the amplitude gain terms in Equation~\eqref{eq::gains}, the phase of the bispectrum, or \emph{closure phase} $\psi$, is preserved under any choice of station-based phase error. The closure phase is a robust interferometric observable: apart from thermal noise, the measured closure phase is the same as the closure phase of the observed image. In the limit of low signal-to-noise, both the bispectrum amplitude and phase are biased by thermal noise and should be debiased before use in imaging \citep{Wirnitzer, Gordon}.

The total number of closure phases at a moment in time is equal to the number of triangles that can be formed from sites in the array, ${N_s\choose3}$. However, not all of these closure phases are independent, as some can be formed by adding or subtracting other closure phases in the set. The total set of independent closure phases can be obtained by selecting an antenna as a reference and choosing only the triangles that include that antenna (\citet{Twiss_60},\citetalias{TMS}). The total number of such independent closure phases is
\begin{equation}
 N_\text{cl phase} = {N_s-1\choose2} =  \frac{(N_s-1)(N_s-2)}{2}.
\end{equation}
$N_\text{cl phase}$ is less than the number of visibilities at a given time, Equation~\eqref{eq::nvis}, by the fraction $1 -2/N_s$. 

Second, on any set of four stations, \emph{closure amplitudes} are formed by taking ratios of visibility amplitudes so as to cancel all the amplitude gain terms in Equation~\eqref{eq::gains}. Up to inverses, the baselines among any set of four stations can form three quadrangles with three corresponding  closure amplitudes
\begin{align}
\label{eq::camps}
 |V_{C}|_a = \left| \frac{V_{12}V_{34}}{V_{13}V_{24}} \right| \;\; &\vphantom{=},\;\;
 |V_{C}|_b = \left| \frac{V_{13}V_{24}}{V_{14}V_{23}} \right|,  \nonumber \\
 |V_{C}|_c &= \left| \frac{V_{14}V_{23}}{V_{12}V_{34}} \right|.
\end{align}
Since the product of the three closure amplitudes in Equation~\eqref{eq::camps} is unity, only two of the set are independent. The total number of closure quadrangles is $3{N_s\choose4}$, but the number of independent closure amplitudes is \citepalias{TMS}
\begin{equation}
 N_\text{cl amp} = \frac{N_s(N_s-3)}{2}.
\end{equation}
$N_\text{cl amp}$ is equal to the total number of visibilities minus the number of unknown station gains. At any given time, the number of closure amplitudes is less than the number of visibilities by a fraction $1 -2/(N_s-1)$. Like the visibility amplitude and bispectrum, closure amplitudes are biased by thermal noise \citepalias{TMS}.

The robustness of closure phases and amplitudes to calibration errors comes with the loss of some information about the source. For instance, closure phases are insensitive to the absolute position of the image centroid, and closure amplitudes are insensitive to the total flux density. These can be constrained separately, either through arbitrary choices (e.g., centering the reconstructed image) or through additional data constraints (e.g., specifying the total image flux density through a separate measurement). 

\subsection{Redundant and Trivial Closure Quantities}
\label{sec::trivial}
Some VLBI arrays include multiple stations that are geographically co-located. For instance, the EHT includes multiple sites on Mauna Kea (the SMA and the JCMT) as well as multiple sites in the Atacama desert in Chile (the ALMA array and the APEX telescope).  Practically, any two sites that form a baseline that does not appreciably resolve any source structure can be considered co-located. 

These ``redundant'' sites can be used to form closure quantities. In the case of closure phase, the added triangles provide no new source information. Specifically, any triangle $\{\vec{b}_{12}, \vec{b}_{23}, \vec{b}_{31}\}$ that includes two co-located sites $\{1, 2\}$ will include one leg that measures the zero-baseline visibility, which has zero phase: $V_{12} = \tilde{I}(0,0)$ is the integrated flux density of the source (see Equation~\ref{eq::VCZ}). The remaining two long legs from the pair of co-located sites to the third site will have $\vec{b}_{23} = -\vec{b}_{31}$ and, consequently, $V_{23} = V_{31}^\ast$. Thus, the bispectrum will be a positive real number, and the closure phase must be zero regardless of the source structure. These \emph{trivial} triangles are not useful for imaging but provide valuable tests of the closure phase statistics and systematic bias \citep[see, e.g.,][]{Fish_2016}. In short, redundant sites can provide additional redundant closure phase triangles, which can be averaged to reduce thermal noise, but they do not give new measurements of source structure.

Redundant sites also give rise to trivial closure amplitudes, which have a value of unity regardless of the source. However, redundant sites also yield new closure amplitudes that are non-trivial and provide additional information on the source structure. For instance, one can measure the normalized visibility amplitude, $\left| V(\vec{u})/V(\vec{0}) \right|$, as a closure quantity on any baseline joining two sets of co-located sites \citep{Johnson_Science}. In the limiting case where every site in an array has a redundant companion, the complete source visibility amplitude information could be recovered through closure amplitudes, except for a single degree of freedom for the total flux density. 

\begin{figure*}[t]
\centering
\includegraphics*[width=.8\textwidth]{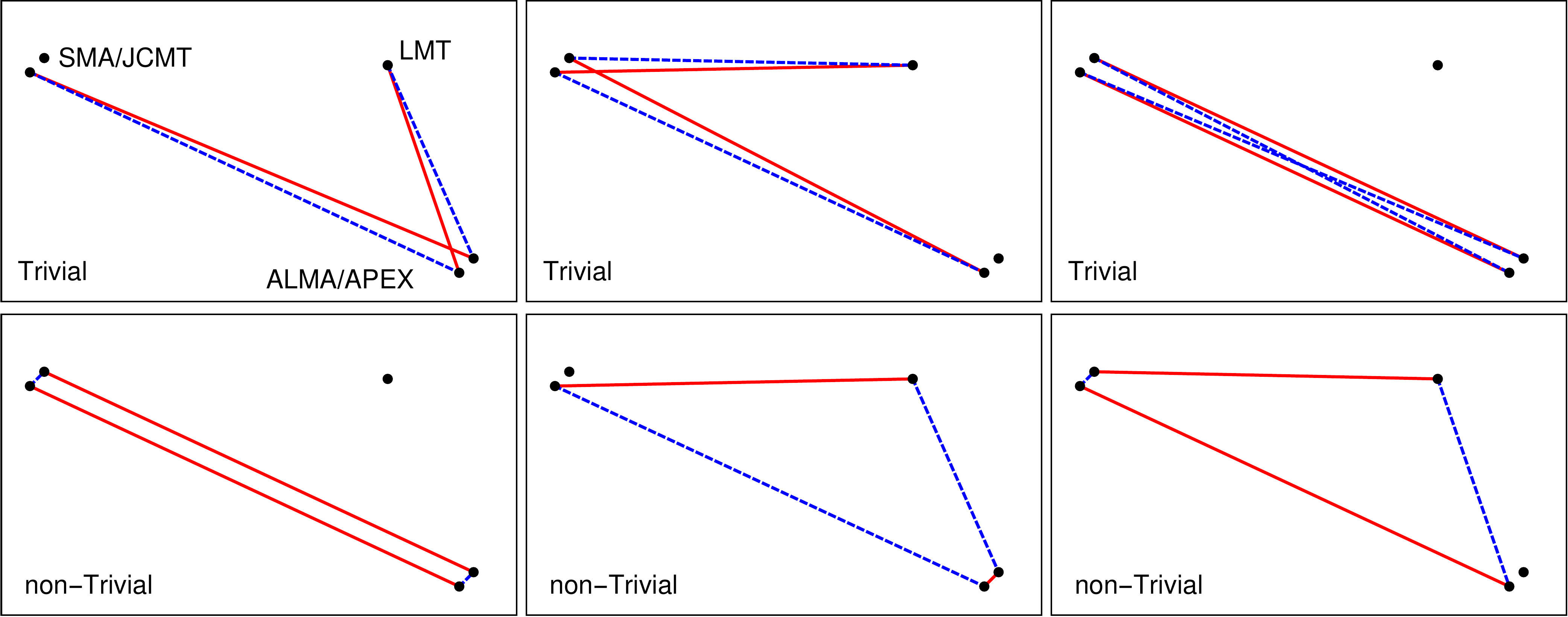} 
\caption{Example closure amplitudes for a portion of the EHT. Solid red lines connecting sites denote visibilities in the numerator of the closure amplitude; dashed blue lines denote visibilities in the denominator. An array containing redundant sites (such as SMA/JCMT and ALMA/APEX in the EHT) will produce new trivial closure amplitudes, which are equal to unity (plus thermal noise), and new non-trivial closure amplitudes, which yield new information about the source. Without redundant sites, there would be no closure amplitudes from this portion of the array.}
\label{fig::trivial_camps}
\end{figure*}

\begin{figure*}[t]
\centering
\includegraphics*[width=.8\textwidth]{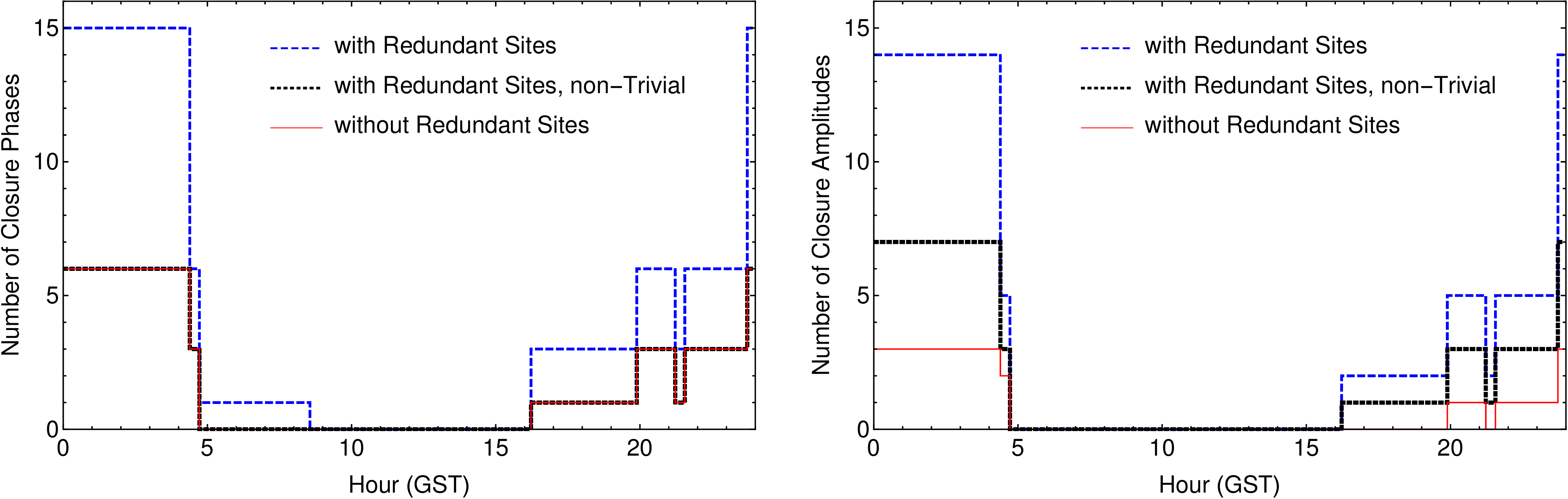} 
\caption{(Left) Number of independent closure phases for the 2017 EHT over 24 hours GMST while observing Sgr A*. The blue line shows the total number of independent closure phases in the array containing redundant stations, the black line shows the number of independent closure phases that measure source structure, and the red line shows the number of independent closure phases in the array when the redundant sites are excluded. Redundant sites do not add any closure phase information to the array apart from decreasing the overall thermal noise. (Right) Total independent (blue) and non-trivial (black) closure amplitudes over 24 hours for the EHT including redundant sites. Unlike for closure phases, adding only two redundant sites significantly increases the amount of information contained in the set of independent closure amplitudes compared to the same array without these sites (red) because not all closure amplitudes containing a baseline between two co-located sites are trivial (see Figure~\ref{fig::trivial_camps}).}
\label{fig::closurecounts}
\end{figure*}

Figure~\ref{fig::trivial_camps} shows examples of the trivial and non-trivial closure amplitudes for an array with partial redundancy. As these examples illustrate, redundant sites can significantly inform and improve calibration and imaging. Figure~\ref{fig::closurecounts} shows the number of closure amplitudes and phases for the EHT with and without redundant sites, both including and excluding trivial additions. The two redundant sites of the 2017 EHT array more than double the amount of information contained in the set of closure amplitudes over the same array without these sites.

\subsection{Thermal Noise on Closure Quantities}
\label{sec::thnoise}
 The thermal noise $\epsilon_{ij}$ on the baseline $i$--$j$ in Equation~\eqref{eq::gains} is a circularly-symmetric complex Gaussian random variable with zero mean that is independently sampled for each visibility measurement. The standard deviation $\sigma_{ij}$ of the thermal noise on this baseline is determined according to the standard radiometer equation \citepalias{TMS}
\begin{equation}
 \label{eq::noise}
 \sigma_{ij} = \frac{1}{\eta} \sqrt{\frac{\text{SEFD}_i \times\ \text{SEFD}_j}{2 \Delta \nu \, \Delta t}}.
\end{equation}
In Equation~\eqref{eq::noise}, $\text{SEFD}_i$ and $\text{SEFD}_j$ are the ``system equivalent flux densities'' of the two telescopes, where for a telescope with system temperature $T_\text{sys}$ and effective area $A_\text{eff}$, the SEFD is $2k_BT_\text{sys}/A_\text{eff}$, with $k_B$ as the Boltzmann constant. The observing bandwidth of the visibility measurement is $\Delta \nu$, and $\Delta t$ is the integration time. The factor of $1/\eta$ in Equation~\eqref{eq::noise} is due to quantization losses in the signal digitization; for 2-bit quantization, $\eta=0.88$ \citepalias{TMS}. 

When the signal-to-noise ratio (SNR) is high, the visibility amplitudes will also be Gaussian distributed with standard deviation $\sigma$ given by Eq.~\eqref{eq::noise}. At lower SNR $>1$, the distribution of the amplitude becomes non-Gaussian, and the estimate of the visibility amplitude taken directly from the norm of the complex visibility is biased upward by the noise. To first order, we debias the amplitudes with the equation \citepalias{TMS}
\begin{equation}
 \label{eq::debias}
 |V|_\text{debiased} = \sqrt{|V|^2_\text{meas} - \sigma^2}.
\end{equation}
For the purposes of this paper, whenever measured visibility amplitudes are used e.g. in the computation of a closure amplitude or $\chi^2$ statistic, we assume they have already been debiased by Equation~\eqref{eq::debias}.  

Turning to the closure quantities, in the high signal-to-noise limit, the baseline-based thermal noise on the closure amplitudes and phases introduced in \S\ref{sec::c_quantities} will also be Gaussian distributed. To first order, the standard deviation of the complex noise on the bispectrum $V_B=V_1 V_2 V_3$ due to the thermal noise on the 3 component visibilities ($\sigma_1$, $\sigma_2$, $\sigma_3$) is 
\begin{equation}
 \label{bierr}
 \sigma_B = |V_B|\sqrt{\frac{\sigma_1^2}{|V_1|^2}+\frac{\sigma_2^2}{|V_2|^2}+\frac{\sigma_3^2}{|V_3|^2}}.
\end{equation}
Then, in the high SNR regime, the standard deviation on the closure phase, $\sigma_{\psi}$ is 
\begin{equation}
\label{eq::cphaseerr}
  \sigma_{\psi} = \frac{\sigma_B}{|V_B|}.
\end{equation}
Similarly, the standard deviation $\sigma_{C}$ of the thermal noise of the closure amplitude $|V_C| =  |V_1 V_2|/|V_3 V_4|$ is, to leading order in the inverse SNR,
\begin{equation}
 \label{eq::camperr}
 \sigma_{C} =  |V_C|\sqrt{\frac{\sigma_1^2}{|V_1|^2}+\frac{\sigma_2^2}{|V_2|^2}+\frac{\sigma_3^2}{|V_3|^2}+\frac{\sigma_4^2}{|V_4|^2}}.
\end{equation}
Note that the expressions for $\sigma_\psi$ and $\sigma_C$ depend only on the measured SNRs of the visibilities and so they too are independent of calibration.

At moderately low SNR, the Gaussianity of the thermal noise on phase and amplitude breaks down, as does the appropriateness of using the measured SNR $|V_i|/\sigma_i$ as an estimate of the true SNR when estimating $\sigma_\psi$ and $\sigma_C$. Because the measured phase is unbiased by thermal noise and wraps at $2\pi$, the true $\sigma_\psi$ is \emph{smaller} than the estimate in Equation~\ref{eq::cphaseerr} in the low-SNR limit. While the chi-squared approach above could be extended to the exact log-likelihood for closure phases with low SNR, low-SNR closure phases are not prone to extreme outliers, so the Gaussian approximation is reasonable to use over a broad range of signal-to-noise. 

The distribution for the reciprocal visibility amplitude, which appears in the denominator of Equation~\eqref{eq::camperr} for $\sigma_C$, takes on an extreme tail at low SNR that extends to positive infinity. This tail causes a large positive bias in the measured closure amplitudes, as well as a severely non-Gaussian distribution. Fitting to log closure amplitudes has the dual benefits of mitigating the tail of the reciprocal amplitude distribution and symmetrizing the numerator and denominator. In this case where the numerator and denominator of the closure amplitude is symmetric, debiasing the component amplitudes with Equation~\eqref{eq::debias} corrects the estimate of the closure amplitude to first order. While detailed analysis of the statistics of closure quantities will be explored in a forthcoming work \citep{lindy_closure}, we have generally found log-closure amplitudes to be more robust observables for imaging (see \S\ref{sec::Results}).  

\section{Imaging with Regularized Maximum Likelihood}
\label{sec::Imaging}

\subsection{Imaging Framework}
\label{sec::imaging}
The standard methods of interferometric imaging are based on the CLEAN algorithm \citep{Hog_1974,Clark_80}. CLEAN operates on the so-called ``dirty image'' obtained by directly taking the Fourier transform of the sparsely sampled visibilities and attempts to deconvolve the ``dirty beam'' which results from the incomplete sampling of the Fourier domain. To perform the initial transform, CLEAN requires well-calibrated complex visibilities. When \emph{a priori} calibration is ineffective, which is often the case at high frequencies when atmospheric phase terms vary rapidly, the visibilities must be ``self-calibrated''. Although self-calibration is used most frequently with CLEAN, it can be used in conjunction with any imaging method that requires calibrated complex visibilities (such as MEM). 

The self-calibration procedure starts from an initial model image and solves for the set of time-dependent complex gains in Equation~\eqref{eq::gains} either by fixing a sufficient set of amplitudes or phases directly from the image and solving for the rest analytically \citep{Wilkinson77, Readhead80}, or by finding a set that minimizes the sum of squares of the differences between the measured and model visibilities \citep{Schwab_1980, Cornwell_1981}. Self-calibration is often performed by first solving only for the phases of the complex gains, correcting the amplitudes at a later stage \citep{Walker_1995,Cornwell_1999}. At each round of self-calibration, the estimated inverse gain terms are then applied to the measured visibilities, and the imager (usually CLEAN) is run again to obtain a new source model. These steps are repeated many times until convergence. There are several assumptions in this procedure which  may affect the final image or the time to convergence of the algorithm. Most critical are the choice of the initial source model (often taken as a point source) and the choice of where to clean the image in each iteration (the so-called ``clean boxes''). These choices enforce assumptions about the source flux distribution early on in the self-calibration process which then propagate to later rounds via the self-calibrated complex visibilities. 

In contrast, the various methods of interferometric imaging explored in this paper all fall under the category of regularized maximum likelihood algorithms. Regularized maximum likelihood methods search for some image that maximizes the sum of a $-\chi^2$ ``data term''  and a ``regularizer'' function that prefers images with certain features when the data are not sufficient to constrain the structure by themselves. These methods can often be interpreted in a Bayesian framework, where the data term is identified with a log-likelihood and the regularizer term with the log-prior. Regularized maximum likelihood methods require only a forward Fourier transform from trial images to the visibility domain. Consequently, they can fit directly to data terms derived from the visibilities even if the visibilities are corrupted by gain and phase errors.  

In astronomy, the most familiar of these methods is the Maximum Entropy Method \citep[see e.g.,][]{Frieden_1972, GS_1978, Cornwell_1985, NN_1986}. While traditional MEM uses calibrated complex visibilities as its fundamental data product, MEM and other, more general regularized maximum likelihood methods have been developed using other regularizers such as the $\ell_1$ norm \citep{Honma_2014} or a Gaussian patch prior \citep{Katie_2015}. Other algorithms have gone beyond complex visibilities as the fundamental data product to produce images directly from the image bispectrum. Development of imaging algorithms that use different fundamental data products from complex visibilities has been particularly fruitful in optical interferometry, where the absolute visibility phase is almost never accessible (e.g. \citet{BSMEM_94, SQUEEZE, Thiebaut_2013, Thiebaut_2017}.), although it has also been explored in the context of VLBI  \citep{Lu_2014, Katie_2015, akiyama_bs}. Regularized maximum likelihood methods have also been extended to polarization \citep{Ponsonby_1973, NN_1983, HW_MEM_1990, Chael_16, couglan_16, akiyama_pol}, to the mitigation of interstellar scattering \citep{stochastic_optics}, and to dynamical imaging to reconstruct movies of time-variable sources \citep{johnson_dyn, starwarps}.

To make our discussion concrete, image reconstruction via regularized maximum likelihood seeks to find an image $\mathbf{I}$ that \emph{minimizes} an objective function $J\left(\textbf{I}\right)$. In this paper, we consider only square images of dimension  $m \times m$. We represent the image $\mathbf{I}$ as a one-dimensional vector of length $M = m^2$. If we consider $N$ observed visibilities, the corresponding sampled Fourier components, or trial image visibilities, $\tilde{V}$ of the trial image vector $\mathbf{I}$ are $\mathbf{V}' = \mathbf{A}\mathbf{I}$, where $\mathbf{A}$ is an $N \times M$ matrix with entries
\begin{equation}
\label{eq::DTFTmatrix}
 A_{ij} = e^{-2\pi \mathrm{i}(u_i x_j + v_i y_j)}.
\end{equation}
Here $(x_j,y_j)$ are the angular coordinates (in radians) of the $j$th pixel, and $(u_i,v_i)$ are the angular frequencies of the $i$th visibility measurement. The direct-time Fourier transform (DTFT) represented by Equation~\eqref{eq::DTFTmatrix} is often the fastest way to compute trial visibilities for sparse arrays observing with narrow fields of view, like the EHT. For large images or large numbers of visibilities, the DTFT is slow and prohibitively expensive in terms of computer memory. In this regime, we must use the Fast Fourier transform (FFT) to obtain the trial visibilities. Algorithms like the Nonequispaced Fast Fourier transform (NFFT, e.g. \citet{NFFT}) are particularly useful for this purpose. In its simplest form, the NFFT takes the FFT of the trial image and interpolates the result to the irregularly sampled $(u,v)$ points. To compensate for inaccuracies in the interpolation procedure, the NFFT both zero-pads the input image and multiplies the pixels by a scaling function (the inverse FT of a convolution kernel in the Fourier domain).

In the most general case, where we may have multiple data terms and multiple regularizers informing our reconstruction, the objective function $J\left(\mathbf{I}\right)$ to be minimized is 
\begin{equation}
 \label{eq::objfunc}
 J(\mathbf{I}) = \sum_{\mathclap{\text{data terms}}} \alpha_D \chi^2_D\left(\mathbf{I},\mathbf{d}\right) - \sum_{\mathclap{\text{regularizers}}} \beta_R S_R\left(\mathbf{I}\right).
\end{equation}
In the above expression, the $\chi^2_D$ are the data terms or chi-squared goodness-of-fit functions corresponding to the data product $\mathbf{d}$. If the data product $\mathbf{d}$ is normally distributed, these are proportional to the log-likelihoods representing the log probability that the data could be observed given an underlying image $\mathbf{I}$. For data products whose distributions are not Gaussian (like closure phases and amplitudes), $\chi^2_D$ is usually an approximation to the likelihood. The $S_R$ are regularizing functions (which we want to \emph{maximize}), which provide missing information on the image characteristics to constrain the space of possible images given our measured data. While relatively new to radio interferometry and VLBI, reconstructions using Equation~\eqref{eq::objfunc} with multiple data terms and regularizers are common in optical interferometry (see e.g. \citet{BSMEM_94,SQUEEZE,Thiebaut_2013,Thiebaut_2017}). 

The ``hyperparameters'' $\alpha_D$ and $\beta_R$ control the relative weighting of the different data and regularizer terms. Because the location of the global minimum of $J(\mathbf{I})$ is unaffected by changes of scale, one hyperparameter can be set to unity or some other arbitrary value without changing the solution. Furthermore, interpreting the $\chi^2_D$ data terms as log-likelihoods, the data term weights $\alpha_D$ should ideally be determined by the number of data points of each type. For example, using the \emph{reduced} $\chi^2$ we define in \S\ref{sec::dterms}, if we set one data term with $N_1$ measurements to $\alpha_1$, the remaining data terms $i>1$ with $N_i$ measurements should all be set as 
\begin{equation}
\label{eq::relweights}
 \alpha_{i>1} = \alpha_i \frac{N_i}{N_1}.
\end{equation}
In practice, we find that with multiple rounds of imaging, heavily weighting a single data term away from the log-likelihood weighting in Equation~\eqref{eq::relweights} can aid initial convergence. We then restore the ideal weighting in Eq.~\eqref{eq::relweights} in later rounds of imaging.

In practice, the hyperparameters $\alpha_D$ and $\beta_R$ are usually adjusted manually to yield reconstructions that converge to the expected values of $\chi^2$ \citep{Cornwell_1985}.  Recently, \citet{akiyama_bs} determined hyperparameters self-consistently using cross-validation. In this method, images are reconstructed with different combinations of the hyperparameters using different data sets where a portion of the data is held in reserve. The set of hyperparameters that produces the image most compatible with the data held in reserve is then used in the final reconstruction.  

\subsection{Data Terms for Robust Imaging}
\label{sec::dterms}
Having defined the general form of the objective function, we turn now to the different choices of the data $\chi^2$ term that can be used in interferometric (total intensity) imaging. The simplest choice is the $\chi^2$ of the measured visibilities $\mathbf{V}$. If there are $N$ total measured visibilities $V_j$, with associated (real) thermal noise RMS values $\sigma_j$, then the reduced $\chi^2$ is
\begin{equation}
 \label{eq::rchisq}
 \chi^2_\text{vis}\left(\mathbf{I}\right) = \frac{1}{2N}\sum_{j}\frac{\left|V_j - V'_j\right|^2}{\sigma^2_j},
\end{equation}
where $V'_{j}$ are the sampled visibilities corresponding to the trial image $\mathbf{I}$.

If the visibility phases are significantly corrupted by atmospheric turbulence, a $\chi^2$ term that uses only the visibility amplitudes can be used: 
\begin{equation}
 \label{eq::rchisqamp}
 \chi^2_\text{amp}\left(\mathbf{I}\right)=\frac{1}{N}\sum_{j}\frac{(|V_j|-|V'_j|)^2}{\sigma^2_j}.
\end{equation}

Because the closure phase is robust to station-based phase errors such as those introduced by atmospheric turbulence, a $\chi^2$ defined on the bispectrum can be used instead of Equation~\eqref{eq::rchisq}. We define $N_B$ as the number of independent bispectrum measurements, and $\sigma_B^2$ as the estimate of the variance on each complex bispectrum measurement (Equation~\ref{bierr}). Then 
\begin{equation}
 \label{eq::chisqb}
 \chi^2_\text{bispec}\left(\mathbf{I}\right)=\frac{1}{2N_B}\sum_j\frac{\left|V_{Bj}-V'_{Bj}\right|^2}{\sigma^2_{Bj}},
\end{equation}
where $V'_{Bj}$ is the sampled bispectrum value corresponding to the trial image $\mathbf{I}$.

We can also define a data term purely using $N_{\psi}$ measured closure phases, $\psi$ (typically $N_{\psi} = N_B$, but we may, e.g., drop trivial closure phases from the fit).  Defining  $\sigma_{\psi}^2$ as their estimated closure phase variances using Equation~\eqref{eq::cphaseerr}, a natural choice of a $\chi^2$ term that automatically respects $2\pi$ phase wraps in the difference of measured and trial image closure phases $\psi$ is    
\begin{align}
\label{eq::chisqcphase}
 \chi^2_\text{cl phase}\left(\mathbf{I}\right)&=\frac{1}{N_{\psi}}\sum_{j}\frac{|e^{\mathrm{i}\psi_{j}}-e^{\mathrm{i}\psi'_{j}}|^2}{\sigma^2_{\psi j}} \nonumber\\
 &=\frac{2}{N_{\psi}}\sum_{j}\frac{1-\cos(\psi_{j}-\psi'_{j})}{\sigma^2_{\psi j}},
\end{align}
where the $\psi'_j$ are the sampled closure phases corresponding to the trial image.

Similarly, a data term that uses only the closure amplitudes $|V_C|$ is
\begin{equation}
\label{eq::chisqcamp}
 \chi^2_\text{cl amp}=\frac{1}{N_{C}}\sum_{j}\frac{(\left|V_{Cj}\right|-\left|V'_{Cj}\right|)^2}{\sigma^2_{Cj}},
\end{equation}
where there are a total of $N_{C}$ measured independent closure amplitudes $|V_{Cj}|$, the $|V'_{Cj}|$ are the corresponding sampled closure amplitudes of the trial image, and the $\sigma^2_{Cj}$ are the estimated variances of the measured closure amplitudes from Equation~\eqref{eq::camperr}.

As discussed in \S\ref{sec::thnoise}, because closure amplitudes are formed from the quotient of visibility amplitudes, the noise on the closure amplitudes (Equation~\ref{eq::camps}) may be highly non-Gaussian. The logarithm  of the closure amplitude will remain approximately Gaussian at lower SNR, so the $\chi^2$ of the logarithm of the closure amplitudes may be a better choice than Equation~\eqref{eq::chisqcamp} in practice. In this case, the $\chi^2$ term is
\begin{equation}
\label{eq::chisqlcamp}
 \chi^2_\text{log cl amp}=\frac{1}{N_{C}}\sum_{j}\frac{\left|V_{Cj}\right|^2}{\sigma^2_{Cj}}\left(\log\frac{\left|V_{Cj}\right|}{\left|V'_{Cj}\right|}\right)^2,
\end{equation}
where we used the fact that to lowest order the variance on the logarithm of a quantity $x$ is $\sigma^2_{\log(x)} = \sigma^2_{x}/x^2$.

\subsection{Data Term Gradients}
\label{sec::gradients}
When using gradient descent algorithms to minimize the objective function (Equation~\ref{eq::objfunc}), providing an analytic expression for the gradient of the objective function with respect to the image pixel values greatly increases the speed  of the algorithm by bypassing the expensive step of estimating gradients numerically. When using a DTFT, the number of computations to evaluate the gradient of a $\chi^2$ term numerically via finite differences is roughly $\mathcal{O}(M^2 \times N)$ (where $M$ is the total number of image pixels and $N$ is the number of measurements). When using an FFT, the scaling is roughly $\mathcal{O}(M \times (M\log M + N))$. In contrast, when using the analytic gradients that we derive below, the corresponding scalings for DTFT and FFT are $\mathcal{O}(M \times N)$ and $\mathcal{O}(M\log M + N)$, respectively. In practice, for typical reconstructions such as those we will show later, analytic gradients improve the imaging speed by a factor comparable to the number of free parameters.

The gradient of the simplest $\chi^2$ term, using complex visibilities (Equation~\ref{eq::rchisq}), is
\begin{equation}
\label{eq::chisqgradvis}
 \frac{\partial}{\partial I_i}\chi^2_\text{vis} = -\frac{1}{N}\sum_{j} \operatorname{Re}\left[A^\dagger_{ij}\left(\frac{V_j-V'_j}{\sigma^2_j}\right)\right].
\end{equation}
The gradients for the other $\chi^2$ terms given in \S\ref{sec::dterms} are presented in the Appendix. 

Note that the visibility $\chi^2$ gradient, Equation~\eqref{eq::chisqgradvis} is the adjoint DTFT of the weighted data residuals. In fact, all the data term gradients considered here can be written as an adjoint DTFT of appropriately weighted residual quantities (see Appendix). 

\subsection{Regularizer Terms}
\label{sec::regs}
To facilitate comparisons across the different data terms in \S\ref{sec::dterms}, we fixed the regularizer terms $S_R$ in the objective function, Equation~\eqref{eq::objfunc}, to be identical for all the reconstructions displayed in this paper. For each reconstruction, we chose to use four regularizer terms.  

The first regularizer is a simple ``entropy'' \citep{Frieden_1972, GS_1978, NN_1986} which rewards pixel-to-pixel similarity to a ``prior image'' with pixel values $P_i$:
\begin{equation}
 \label{eq::entropy}
 S_\text{entropy} = -\sum_{i=1}^n I_i \log\left(\frac{I_i}{P_i}\right).
\end{equation}
For the second regularizer, we use one of two forms of a ``smoothness'' constraint that pushes the final image to favor pixel-to-pixel smoothness. The first is an isotropic total variation regularizer (or $\ell_2$ norm on the  image gradient) that favors piecewise-smooth images with flat regions separated by sharp edges \citep{TV}.
\begin{equation}
\label{eq::tv}
 S_{TV} = -\sum_{l}\sum_{m} \left[ \left(I_{l+1,m}-I_{l,m}\right)^2 + \left(I_{l,m+1}-I_{l,m}\right)^2\right]^{1/2},
\end{equation}
where in the above equation the two sums  are taken over the two image dimensions and the image pixels $I_{l,m}$ are now indexed by their position $(l,m)$ in the 2D $m \times m$ grid. It should be noted that the total variation in Equation~\eqref{eq::tv} is not everywhere differentiable, so care must be taken when using it in imaging. \citet{Thiebaut_2017} presents a differentiable hyperbolic form of an edge-preserving smoothness regularizer \citep{Charb1997} which approximates TV when the image is far from being smooth ($S_{TV}$ is large).  

In the reconstructions in \S\ref{sec::EHT}, however, we instead use a ``Total Squared Variation'' regularizer which favors smooth edges and may be more appropriate for astronomical image reconstruction \citep[see forthcoming work by][]{kuramochi_tv2}.
\begin{equation}
\label{eq::tsv}
 S_{TSV} = -\sum_{l}\sum_{m} \left[ \left(I_{l+1,m}-I_{l,m}\right)^2 + \left(I_{l,m+1}-I_{l,m}\right)^2\right],
\end{equation}

The third and fourth regularizers constrain image-averaged properties. First, because closure amplitudes are independent of the normalization of the image, we include a constraint on the total image flux density:
\begin{equation}
 S_\text{tot flux} = -\left(\sum_i I_i - F\right)^2,
\end{equation}
where the sum is over the $M$ pixels in the image and $F$ is the total source flux density, considered to be known \emph{a priori} (e.g. by a simultaneous measurement of the source by a flux-calibrated single station). Next, because closure phase does not constrain the position of the image centroid, we also include a regularizing constraint to center the image in the chosen field of view: 
\begin{equation}
 S_\text{centroid} = -\left(\sum_{i}I_ix_i - F\delta_x\right)^2 + \left(\sum_{i}I_iy_i - F\delta_y\right)^2,
\end{equation}
where $(x_i,y_i)$ is the coordinate of the $i$th pixel and the desired image centroid position is $(\delta_x,\delta_y)$. In this paper, we use coordinates where $(x_i,y_i) = (0,0)$ is in the center of the frame and set $(\delta_x,\delta_y) = (0,0)$. When only closure phases and closure amplitudes are used in the reconstruction, both the centroid and the total flux density are completely unconstrained by data. Thus, in this case almost any amount of weight on $S_\text{tot flux}$ and $S_\text{centroid}$ should guide the final image to a centered image with the specified total flux, and the precise weighting of these terms relative to the data is not as significant in informing the final image as the relative weighting of $S_{TV}$ or $S_{TSV}$ and $S_\text{entropy}$. 

While the four above regularizers are used for all of the data sets imaged in this paper, we adjusted their relative weighting (the $\beta_R$ terms in Equation~\eqref{eq::objfunc}) as well as the prior image used in Equation~\eqref{eq::entropy} based on the data set considered. However, when comparing images produced with different data terms from \S\ref{sec::dterms}, we were consistent in using the same prior image and relative regularizer weightings in the different reconstructions to produce fair comparisons. Furthermore, when imaging synthetic datasets in \S\ref{sec::EHT}, we used the same combination of regularizer weights (see Table\ref{tab::eht_params}). The prior images in Equation~\ref{eq::entropy} are different from dataset to dataset in this section, but in every case we use a relatively uninformative prior consisting of a Gaussian with a size of roughly half of the reconstruction field of view, and the weighting of Equation~\ref{eq::entropy} is small. 

\section{Implementation}
\label{sec::imp}
\begin{table*}[!t]
\begin{center}
\caption{EHT 2017 Station Parameters}
\label{tab::eht_station}
\begin{tabular}{ll|ll|lll}
\hline \hline
Facility & Location & Diameter (m)  & $\mbox{SEFD}$ (Jy) & X (m) & Y (m)& Z (m)\\
\hline
JCMT     & Mauna Kea, USA& 15           & 6000  &  -5464584.68 & -2493001.17  &  2150653.98 \\
SMA      & Mauna Kea, USA& $7(\times6)$ & 4900 &  -5464555.49 &  -2492927.99 &  2150797.18 \\
SMT      & Arizona, USA& 10           & 5000   &  -1828796.2 &  -5054406.8 &  3427865.2 \\
APEX     & Atacama Desert, Chile & 12           & 3500   & 2225039.53  &  -5441197.63 &  -2479303.36 \\
ALMA     & Atacama Desert, Chile &$40(\times12)$ & 90 &  2225061.164 & -5440057.37  &  -2481681.15 \\
SPT   & South Pole & 10             & 5000 &  0.01 &  0.01 &  -6359609.7 \\
LMT     & Sierra Negra, Mexico & 50             & 600  &   -768715.63&   -5988507.07&  2063354.85 \\
IRAM      &Pico Veleta, Spain & 30              & 1400 &  5088967.75 &  -301681.186 & 3825012.206  \\
\end{tabular}\\
\end{center}
\bigskip
Current EHT sites are the Atacama Large Millimeter/Submillimeter Array (ALMA), the Large Millimeter Telescope (LMT), the Submillimeter Array (SMA), the Submillimeter
Telescope (SMT), the Institut de Radioastronomie Millim\'{e}trique (IRAM) telescope on Pico Veleta (PV), the IRAM Plateau de Bure
Interferometer (PdB), and the South Pole Telescope (SPT). Note
that PdB did not participate in 2017 EHT observations.
\end{table*}

\begin{figure*}[hbtp]
\centering
\includegraphics*[width=.8\textwidth]{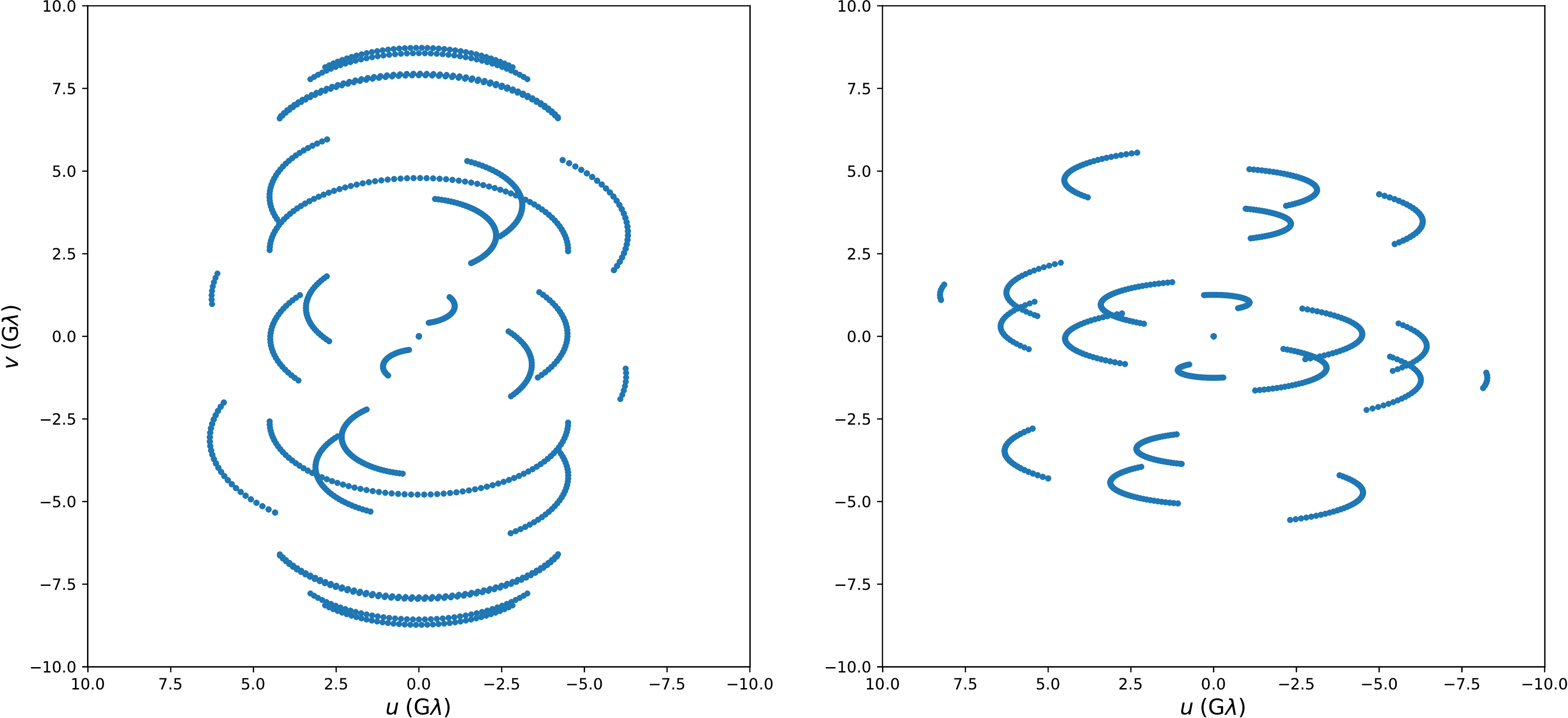} 
\caption
 {(Left) EHT 2017 $u,v$ coverage for \sgra. The ``redundant'' (JCMT and APEX) sites make practically no unique contributions to the $u,v$ coverage or nominal resolution aside from adding an effective zero baseline. However, these sites add closure amplitudes that are essential for closure amplitude imaging to approach the fidelity of imaging with visibility amplitudes. (Right) EHT 2017 $u,v$ coverage for M87.
 }
\label{fig::uv}
\end{figure*}

\subsection{Imaging Methods}
\label{sec::implementation} 
We implemented the imaging framework described in \S\ref{sec::Imaging} including 
all of the data terms introduced in \S\ref{sec::dterms} in the 
\texttt{eht-imaging} software library \citep{ehtim}, originally developed for 
polarimetric VLBI imaging \citep{Chael_16}. To minimize the objective function, 
Equation~\eqref{eq::objfunc}, using different combinations of data terms and 
regularizers, the imaging routines in \texttt{eht-imaging} use the 
Limited-Memory BFGS algorithm \citep{LBFGS} as implemented in  the 
\texttt{Scipy} package \citep{SCIPY}. The L-BFGS algorithm is a quasi-Newton 
gradient descent method that uses the analytic forms of the data term gradients 
presented in \S\ref{sec::gradients} to progress toward a minimum in the 
objective function. 

As described in \citet{Chael_16}, our imaging algorithm ensures a positive flux in each pixel by performing a change of variables $I_i = \exp{\xi_i}$, where $-\infty<\xi_i<\infty$. When imaging in the log intensity domain, the gradients in \S\ref{sec::gradients} must be multiplied by $\exp{\xi_i}$. We also use the continuous image representation introduced in \citet{Katie_2015} where each array of pixel intensities is taken to represent a continuous function formed by convolving a comb of Dirac delta functions with a pixel ``pulse'' function. Introducing a continuous image representation multiplies the visibilities of the discrete image array by a taper given by the Fourier transform of the pulse function, removing spurious high-frequency structure introduced by the regular pixel spacing. For this paper, we used a triangular pulse function in both dimensions with width $2\Delta$, where $\Delta$ is the image pixel spacing. 

Finally, to aid in convergence and help the minimizer avoid local minima in the objective function, we run each imager multiple times for each dataset, substituting a blurred version produced by convolving the result of the previous run with a circular Gaussian as the next initial image. This procedure smooths out initial spurious high-frequency artifacts that the imager will not remove on its own given a lack of data constraints. Each time we restart the imager we also adjust the various hyperparameters $\alpha_D$ and $\beta_R$ in Equation~\eqref{eq::objfunc}. The prescriptions for each data set are presented below, but in general our approach is to generally increase the weight on the smoothness regularizer term to suppress the emergence of spurious high-frequency artifacts. We also usually begin by weighting the closure phase data term more heavily in the reconstruction than is supported by the log-likelihood interpretation (Equation~\eqref{eq::relweights}), since we find that minimizing the closure phase $\chi^2$ is more helpful in constraining the overall image structure in early rounds of imaging. As we progress to later rounds, we restore the relative data term weighting to that given by Equation~\eqref{eq::relweights}. 

On small data sets or on reconstructions with a small field of view, as is the case for simulated EHT data (\S\ref{sec::EHT}), direct-time Fourier transforms are sufficient to compute the data terms in \S\ref{sec::dterms}. For larger data sets such as those produced by the VLBA and especially ALMA (\S\ref{sec::ALMA}), the DTFT matrix $A_{ij}$ becomes prohibitively large to store in memory and prohibitively slow at extracting visibilities from the trial image at each step. In this regime, we use the Nonequispaced Fast Fourier Transform package (NFFT \citet{NFFT}) accessible in Python via the \texttt{pyNFFT} wrapper.\footnote{\url{https://pypi.python.org/pypi/pyNFFT}} The forward NFFT is used for the computation of the irregularly sampled visibilities from the regularly sampled input image, and the adjoint NFFT is used to compute image domain gradient components from data term residuals (see the Appendix). 

\subsection{Simulated Data}
\label{sec::simdata}

To test the effects of using the different data terms in \S\ref{sec::dterms} in VLBI imaging, we simulated VLBI observations from model images and applied different amounts of uncertainty in the complex station gains $G_i e^{\mathrm{i}\phi_i}$ (Equation~\eqref{eq::gains}). We then produced images using different data term combinations and the regularizers as described in \S\ref{sec::implementation} and compared our results with the true source images. 

To generate synthetic data with different degrees of gain error, we generated time-dependent station-based complex gains sampled from known underlying distributions. Because the atmospheric coherence time which determines the additional phase $\phi_i$ added at each station is much shorter than a typical observing cadence at 1.3 mm, we sampled these phases from a uniform distribution over $-\pi < \phi_i < \pi$ at each time, independent of the uncertainty in the amplitude. 

Our prescription for the amplitude gain terms consisted of a random time-independent offset and a fluctuating part:
\begin{equation}
 |G_i| = \sqrt{(1+X_i)(1+Y_i(t))},
\end{equation}
where $X_i$ and $Y_i$ are real Gaussian random variables with zero mean, but $X_i$ is drawn only once per telescope per observation and $Y_i$ is drawn independently at each time when $u,v$ points are sampled. For simplicity, we chose to use identical standard deviations for the underlying Gaussian distributions of $X_i$ and $Y_i$ and call this standard deviation our level of gain error. 

We applied our different sampled sets of station based gains computed at different levels of amplitude error to the ideal visibilities plus identical Gaussian thermal noise using Equation~\eqref{eq::gains}. To preserve the signal-to-noise ratio, the reported noise standard deviation terms $\sigma_{ij}$ from Equation~\eqref{eq::noise} were multiplied by the same gain factors $G_i$, and $G_j$. 

We also included the effects of varying elevation and opacity $\tau$ on our signal-to-noise at each site. The opacity attenuates the measured perfect visibility $V_{ij}$ (before adding thermal noise) by a factor $\sqrt{e^{-\tau_i/\sin\theta_i}e^{-\tau_j/\sin\theta_j}}$, where $\theta_i$ and $\theta_j$ are the elevation angles of the source at the different telescopes. This reduces the signal-to-noise by the same factor $\sqrt{e^{-\tau_i/\sin\theta_i}e^{-\tau_j/\sin\theta_j}}$. This factor can be corrected for by multiplying the measured visibility (including thermal noise) by its inverse using the \emph{measured} opacity, keeping the reduced signal-to-noise constant. In general, the imperfect measurement of opacities introduces an additional source of amplitude gain error. For the purposes of this paper, when we simulate data we assume the perfect measurement of opacities and set all zenith opacities $\tau_i = 0.15$. 

\subsection{Image Evaluation}
\label{sec::eval}
To evaluate the fidelity of images reconstructed from data, we followed \citet{Chael_16} and \citet{akiyama_bs} in using a simple normalized root-mean-square error (NRMSE) fidelity metric. The NRMSE is a point-to-point metric that evaluates images based on pixel-to-pixel similarities rather than common large-scale features. Given two images $\mathbf{A}$ and $\mathbf{B}$ with $M$ pixels each, the NRMSE of image $\mathbf{A}$ relative to $\mathbf{B}$ is
\begin{equation}
\label{eq::ncc}
 \text{NRMSE}(\mathbf{A},\mathbf{B}) = \frac{\sqrt{\sum_{i=1}^{M}(A_i - B_i)^2}}{{\sqrt{\sum_{i=1}^{M}{B_i^2}}}},
\end{equation}

\begin{table*}[!t]
\centering
\caption{Initial/Prior image parameters}
\label{tab::eht_params2}
\begin{tabular}{l||lllll}
\hline \hline
Image & $u,v$ coverage & FOV ($\mu as$)  & Gaussian FWHM & Flux (Jy)($\mu$as) \\
\hline
Figure~\ref{fig::eht_dataset_2} & \sgra  &  135  & 60&2 \\
Figure~\ref{fig::eht_dataset_5} & M87 &  155  & 60&2\\
Figure~\ref{fig::eht_dataset_1} & \sgra  &  255  & 80&2\\
Figure~\ref{fig::eht_dataset_3} & \sgra &  375  & 25 &2
\end{tabular}
\end{table*}

\begin{table*}[!t]
\centering
\caption{Imaging Parameters}
\label{tab::eht_params}
\begin{tabular}{l||l|llll|ll|ll}
\hline \hline
Round & $f_\text{blur}$ & $\beta_\text{entropy} $ & $\beta_{TSV}$ & $\beta_\text{tot flux}$ & $\beta_\text{centroid}$& $\alpha_1$ & $f_{\alpha 2}$ & $N_\text{iter}$ & $\epsilon_\text{conv}$ \\
\hline
1 & N/A  &  1 & 1   & 100 & 100 & 100 & 2    & 50  & $10^{-10}$ \\
2 & 0.75 &  1 & 50  & 50  & 50  & 100 & 0.75 & 150 & $10^{-10}$ \\
3 & 0.5  &  1 & 100 & 10  & 10  & 100 & 0.5  & 200 & $10^{-10}$ \\
4 & 0.33 &  1 & 500 & 1   & 1   & 100 & 1    & 200 & $10^{-10}$ 
\end{tabular}
\end{table*}

Several factors complicate the simple application of Equation~\eqref{eq::ncc} in evaluating our reconstructed images. First, often the true source image will contain fine-scale features that are at too high a resolution for any image reconstruction algorithm to capture given the longest projected baseline in the $u,v$ plane. To prevent NRMSE from unduly penalizing reconstructions that successfully reconstruct the lower resolution features in the data, we convolve both the true and reconstructed image with a Gaussian kernel to blur out high-frequency structure. Since we expect our algorithms to provide some ``super-resolution'' above the scale corresponding to the longest projected baseline, we choose to blur the images with a Gaussian that has the same proportions as the interferometer ``clean'' beam -- the Gaussian fitted to the central lobe of the Fourier transform of the $u,v$ coverage -- but we scale the beam size by a factor of 1/3. 

A second complication arises because images reconstructed without calibrated visibility phases are not  sensitive to the true position of the image centroid in the field of view, so reconstructed images may be offset  from the true source location. In addition, the number of pixels and field of view in the reconstructed image may be different from those in the true source image. Therefore, when comparing images we first resample our reconstructions onto the same grid as the model image using cubic spline interpolation and then find the overall centroid shift of the reconstruction that produces the maximal cross-correlation before computing the NRMSE with Equation~\eqref{eq::ncc}.

\section{Results}
\label{sec::Results}
\subsection{Results: Simulated EHT Images}
\label{sec::EHT}

We simulated data on EHT baselines from several 230 GHz model images at the positions of the EHT's primary science targets: \sgra\ (RA: 17h 45m 40.04s, DEC: $-29^\circ$ $0'$ $28.12''$) and M87 (RA: $12^\text{h}$ $30^\text{m}$ $49^\text{s}.42$, DEC: $+12^\circ$ $23'$ $28.04''$). Our model images were generated by performing General Relativistic ray tracing and radiative transfer on the density and temperature distributions from GRMHD simulations of hot supermassive black hole accretion disks \citep{Chan15, Mosc16, Gold_Pol}. We also simulated data from a 7mm VLBA image of the quasar 3C273 \citep{Jorstad16} rescaled to a smaller FOV of $250\mu$as, which we placed at the sky location of \sgra. 

The EHT's station parameters are listed in Table~\ref{tab::eht_station}. In addition to the full EHT array described in Table~\ref{tab::eht_station}, we also generated data on a reduced array without the ``redundant'' sites -- JCMT and APEX -- that are located at the same location as the more sensitive SMA and ALMA, respectively. The $u,v$ coverage maps for the 2017 EHT when observing \sgra\ and M87 are displayed in Figure~\ref{fig::uv}.

In all cases, we use an integration time $\Delta t=30$s and a bandwidth $\Delta \nu = 2$GHz, with scans taken every 5 minutes for a full 24 hour rotation of the Earth. The zenith opacity was set to $\tau=0.15$ at all sites with no uncertainties in the opacity calibration. We did not include the effects of either refractive or diffractive interstellar scattering in our simulated \sgra\ data (see e.g. \citet{Fish_2014, stochastic_optics}). 

To test the quality of our different imaging methods with different levels of gain uncertainty, we produced one dataset for each image with only thermal noise, then generated random gain terms at seven different levels of uncertainty -- 0\%, 5\%, 10\%, 25\%, 50\%, 75\%, and 100\% -- as described in \S\ref{sec::simdata}. 

We reconstructed each dataset on a $128\times128$ pixel grid using each of four different data term combinations: bispectrum, (Equation~\ref{eq::chisqb}), visibility amplitude and closure phase (Equation~\ref{eq::rchisqamp} and~\ref{eq::chisqcphase}), closure amplitude and closure phase (Equation~\ref{eq::chisqcamp} and~\ref{eq::chisqcphase}) and log closure amplitude and closure phase (Equation~\ref{eq::chisqlcamp} and~\ref{eq::chisqcphase}). In all our reconstructions, we used all four regularizer terms introduced in \S\ref{sec::regs} (using ``Total Squared Variation'' as our smoothness regularizer, Equation~\eqref{eq::tsv}). To ensure consistency in our comparisons, we followed the same imaging procedure for all images, arrays, and methods. For each dataset, we only changed the image field of view and corresponding initial image, which is also used as the prior in the $S_\text{entropy}$ regularizer. The initial/prior image was a circular Gaussian in all cases. The total fluxes, fields of view, and initial image FWHMs are given in Table~\ref{tab::eht_params2}. 

The parameters that specify our imaging procedure are listed in Table~\ref{tab::eht_params}. As mentioned in \S\ref{sec::implementation}, we image each dataset in multiple rounds, blurring out the final image from a given round to serve as the initial image in the next. The FWHM of the circular Gaussian blurring kernel used is given as a fraction $f_\text{blur}$ of the nominal array resolution.  The other imaging parameters listed in Table~\ref{tab::eht_params} include the data term and regularizer hyperparameters, $\alpha_D$ and $\beta_R$. For the data terms, in each case $\alpha_1$ refers to the amplitude term (bispectrum, visibility amplitude, closure amplitude, or log closure amplitude), and $\alpha_2$ is the hyperparameter for the closure phase term, if present (all methods except bispectrum). We parametrize  $\alpha_2$ by stating its ratio $f_{\alpha 2}$ with the correct log-likelihood ratio given by Equation~\eqref{eq::relweights}. That is, if there are $N_1$ measurements of the first (amplitude) data product and $N_2$ measurements of of the second (phase) data product, 
\begin{equation}
 \label{eq::relweights2}
 \alpha_2 = f_{\alpha 2} \,\alpha_1 \,\frac{N_2}{N_1}.
\end{equation}

Finally, we also list the maximum number of imager steps allowed in each round, $N_\text{iter}$, and the convergence criterion $\epsilon_\text{conv}$ for the minimum allowed fractional change in the objective function and gradient magnitude between imager steps. 

\begin{figure*}[htbp]
 \centering
 \includegraphics*[width=\textwidth]{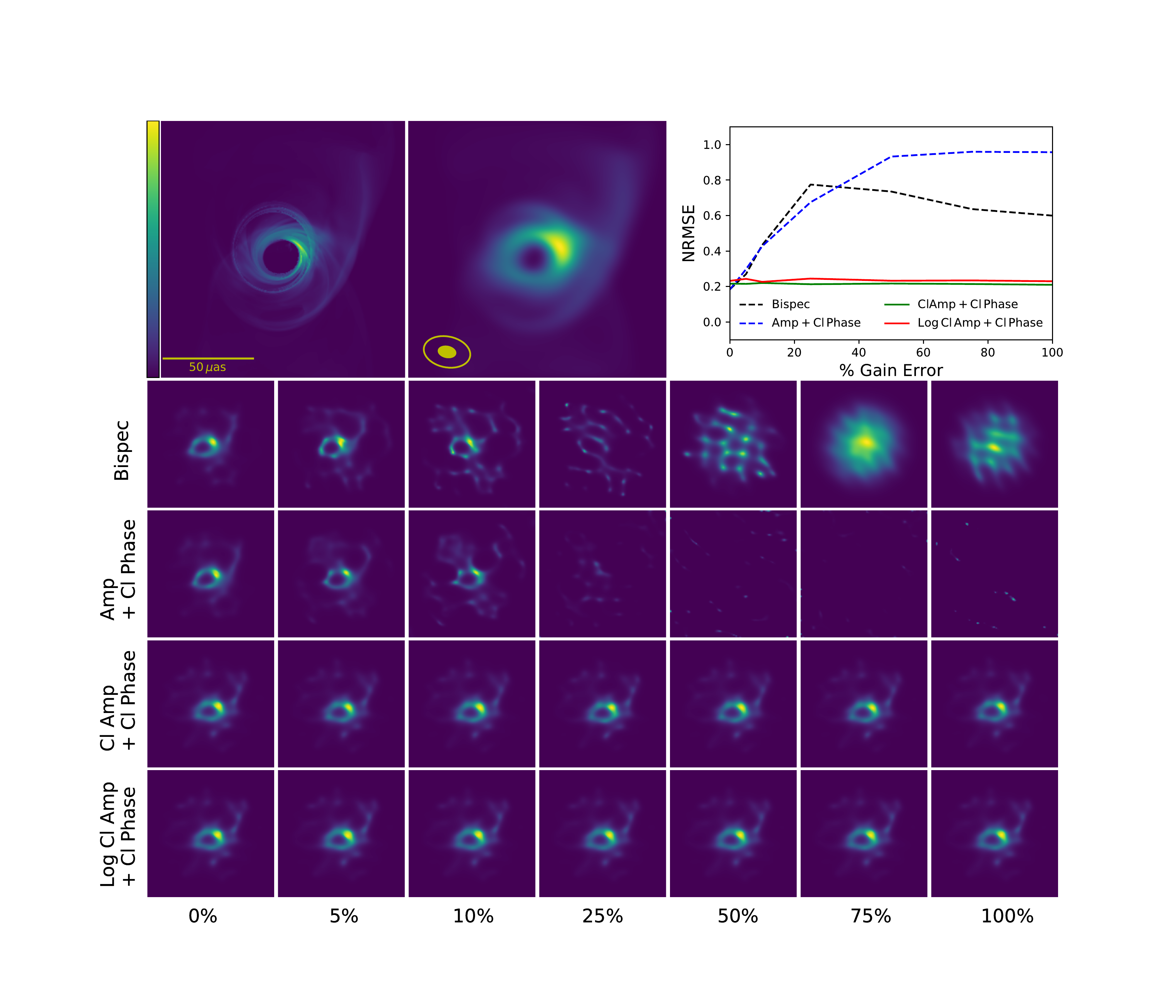}
 \caption
 {(Top left) 230 GHz image from a GRMHD simulation of \sgra\ \citep{Gold_Pol}. (Top middle) the same image blurred with the effective beam (solid ellipse), 1/3 the size of the fitted CLEAN beam (open ellipse). The image was observed at the sky location of \sgra\ using EHT 2017 baselines, and images were reconstructed with each method using the parameters in Table~\ref{tab::eht_params}. (Top right) Curves of NRMSE (Equation~\ref{eq::ncc}) versus gain error for each  reconstruction method. (Bottom) individual reconstructions from each method (y-axis) at each level of gain error (x-axis), blurred with the same beam as the model in the upper middle pane.  The images and NRMSE curves show that except at the lowest levels of amplitude gain error, the closure-only results are as faithful to the model as the reconstructions that use either the bispectrum or visibility amplitudes and closure phases. Furthermore, the results of the closure-only methods are insensitive to the level of amplitude gain error, while the reconstructions using visibility amplitude information fail completely starting at the 25\% level of gain error. 
 }
\label{fig::eht_dataset_2}
\end{figure*}

\begin{figure*}[htbp]
 \centering
 \includegraphics*[width=\textwidth]{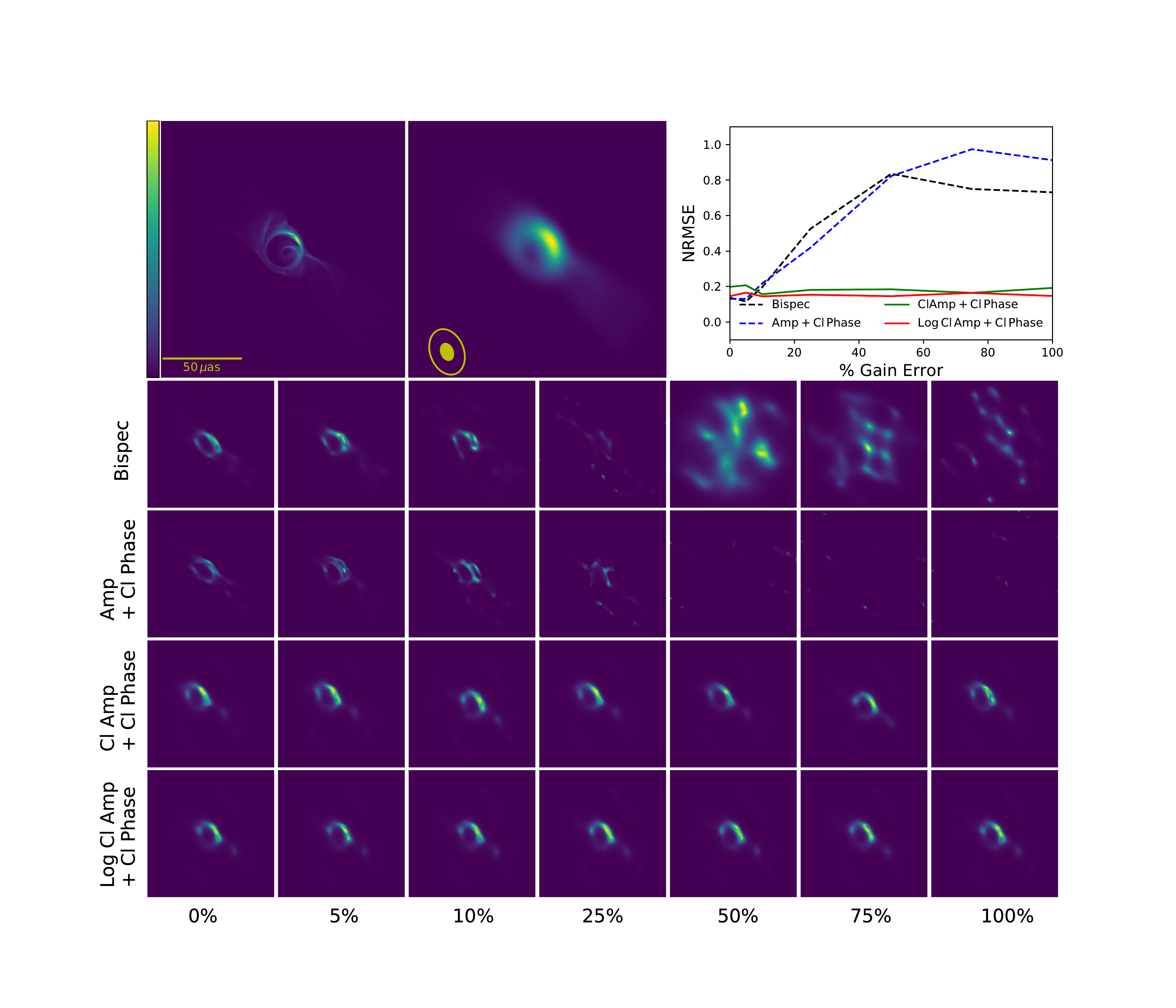}
 \caption
 {Reconstructions of a 230 GHz image from a GRMHD simulation of the M87 jet \citep{Mosc16}. As in the \sgra\ image in Figure~\ref{fig::eht_dataset_2}, closure-only methods produce results that are as good or better than the bispectrum or visibility amplitude + closure phase methods in all but the zero gain error case,  and the closure-only results are consistent at all levels of gain error.  
 }
\label{fig::eht_dataset_5}
\end{figure*}

\begin{figure*}[htbp]
 \centering
 \includegraphics*[width=\textwidth]{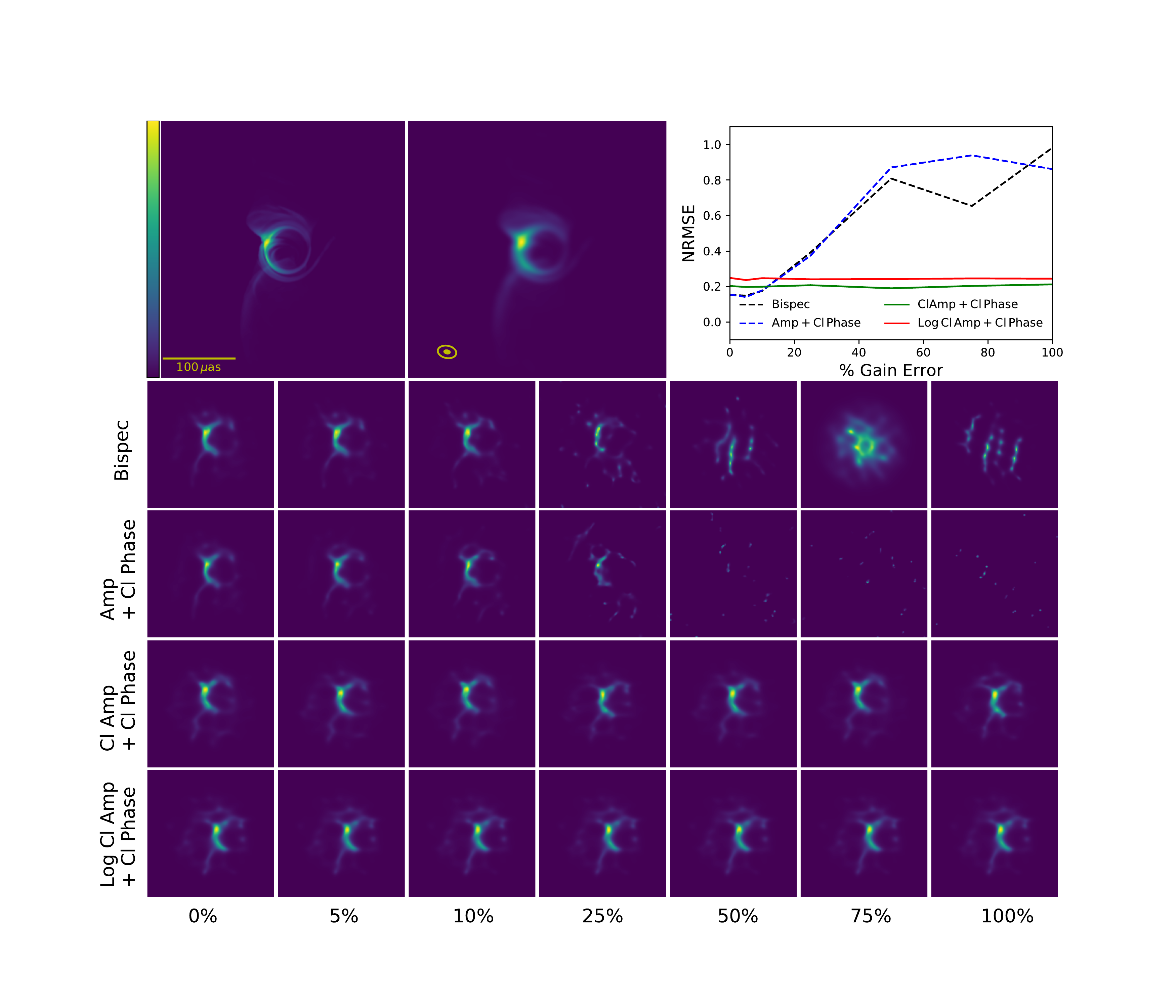}
 \caption
 {Reconstructions of a 230 GHz image from a GRMHD simulation of the \sgra\ accretion flow \citep{Chan15}. Both the closure amplitude and log closure amplitude reconstructions performed consistently at all levels of gain error.
 }
\label{fig::eht_dataset_1}
\end{figure*}

\begin{figure*}[htbp]
 \centering
 \includegraphics*[width=\textwidth]{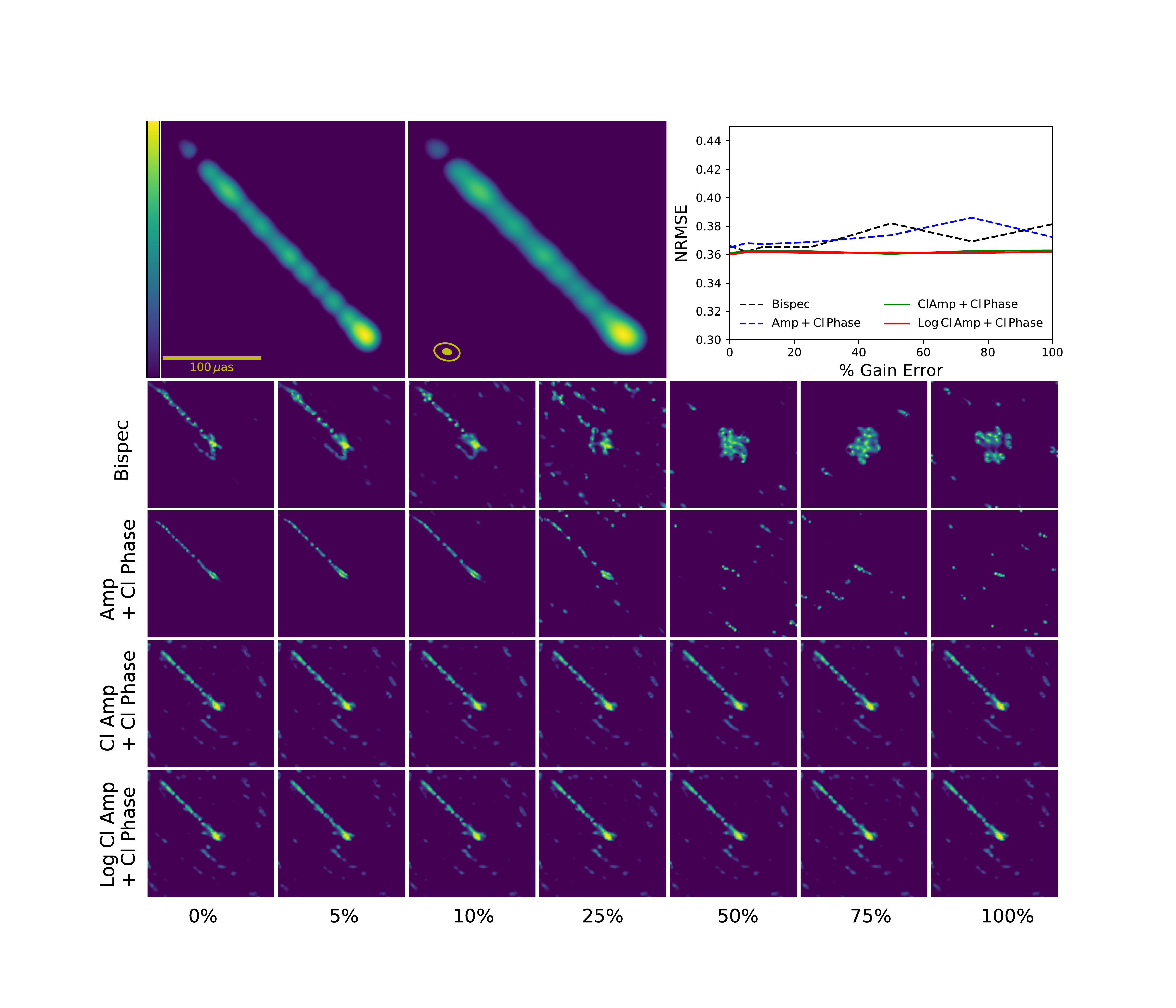}
 \caption
 {
 43 GHz VLBA image of 3C273 from \citet{Jorstad16}, scaled to a 250 $\mu$as field of view. Simulated data were generated using the 230 GHz EHT 2017 \sgra\ $u,v$ coordinates and sensitivities (Figure~\ref{fig::uv}). Unlike the other images in this section, this image is displayed with a log scale, and the NRMSE was computed from the log of the image. The closure-only reconstructions again capture the overall jet structure at all levels of amplitude gain error. With no gain error, imaging directly with closure amplitudes (or log closure amplitudes) instead of visibility amplitudes provides less dynamic range, as is evident from the spurious low-luminosity off-axis features in the closure-only reconstructions, likely resulting from a local minimum in  the objective function. }
\label{fig::eht_dataset_3}
\end{figure*}

Our results are displayed in Figs.~\ref{fig::eht_dataset_2}, \ref{fig::eht_dataset_5}, \ref{fig::eht_dataset_1}, and \ref{fig::eht_dataset_3}. In each figure we show the initial model image, the initial model image blurred with a ``clean'' beam scaled to 1/3 of its fitted value, and the reconstructions from each method for each level of gain uncertainty, all blurred with the same beam. We also display a plot showing the normalized root-mean-square error (Equation~\ref{eq::ncc}) for each method as a function of the level of gain error in the underlying dataset. 

Our results indicate that, as long as some redundant sites are included to constrain the reconstruction with ``trivial'' closure phases and amplitudes, closure-only imaging of EHT data can achieve fidelities nearly as good as bispectral or amplitude + closure phase imaging. As the level of amplitude gain error increases, the fidelity of the results produced using the bispectrum or visibility amplitudes drops quickly,  while closure-only imaging is completely insensitive to gain error. 

Figures~\ref{fig::eht_dataset_2}, \ref{fig::eht_dataset_5}, and \ref{fig::eht_dataset_1} show that imaging with closure amplitudes directly can produce results that are more faithful to the underlying image than reconstructing the image with log closure amplitudes. However, we have found imaging with the closure amplitudes often takes much longer to converge, and is more sensitive to the particular choices of data term weight and initial field of view.



Finally, for the narrow, high dynamic range scaled 3C279 image in Figure~\ref{fig::eht_dataset_3}, we computed the NRMSE using the logarithm of the image. This results in a range of NRMSE values for the bispectrum and visibility amplitude + closure phase images that is substantially lower than those in Figures~\ref{fig::eht_dataset_2},~\ref{fig::eht_dataset_5}, and~\ref{fig::eht_dataset_1}; however, visual inspection of the images shows that in this case, as in Figures~\ref{fig::eht_dataset_2} and~\ref{fig::eht_dataset_5}, imaging methods that rely on calibrated amplitudes perform significantly worse with increasing gain error and completely fail with amplitude gain error levels $> 25$\%. In contrast, the closure-only methods have consistent performance at all levels of amplitude gain. However, the final dynamic range achieved in the closure-only reconstructions is worse than in the images produced with visibility amplitudes with zero gain error, as is evident from the spurious low-luminosity features in the closure-only reconstructions in Figure~\ref{fig::eht_dataset_3}. These features parallel to the jet axis result from being trapped in a local minimum of the objective function, which is invariant to overall image shifts. Since there are no data constraints on certain spatial frequencies due to sparse coverage, these Fourier components can be made large through periodic structure without increasing $\chi^2$. Defining a masked region along the jet axis outside which the flux is zero (analogous to a CLEAN box) may help remove these features. 

\begin{figure}[t!]
\centering
 \includegraphics*[width=\linewidth,trim={.0005cm 0.0005cm .0005cm .0005cm},clip]{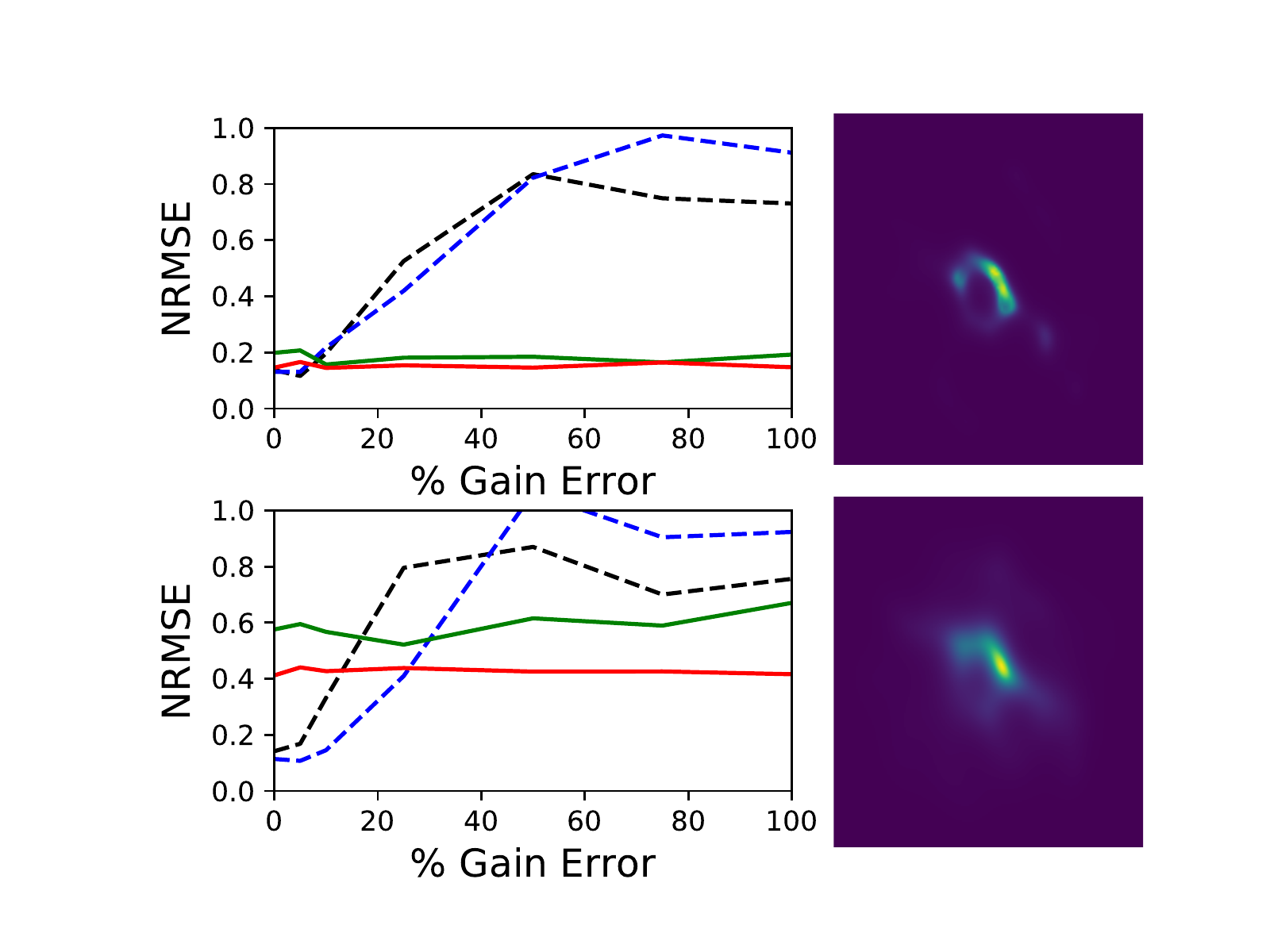}
\caption
 {(Top) Image fidelity with the EHT 2017 array. The left panel shows NRMSE curves of image fidelity for reconstructions of the model in Figure~\ref{fig::eht_dataset_2} with  different levels of gain error. The curves are styled consistently with those in Figures~\ref{fig::eht_dataset_2}--\ref{fig::eht_dataset_3}. The right image is the log closure amplitude + closure phase reconstruction produced at 100\% gain error. (Bottom) Image fidelity with the EHT 2017 array without redundant stations (JCMT, APEX). The reconstructions from data without including the redundant stations are still insensitive to different levels of gain error, but their overall fidelity is worse compared with those produced from data including these redundant stations.}
\label{fig::redundant_compare}
\end{figure}

We also compared reconstructions using data from the full EHT 2017 array and the 2017 array without ``redundant'' sites. Figure~\ref{fig::redundant_compare} shows that in both cases closure-only methods converge to the same reconstruction for all values of systematic gain error. However, without redundant sites the results are substantially less accurate, while using a redundant sites in the dataset, the closure-only results approach the fidelity of images produced with gain-calibrated amplitudes.  ``Redundant'' sites contribute important short baselines that combine into nontrivial closure amplitudes and act to further constrain the underlying image (\S\ref{sec::trivial}). In other words, the closure-only images approach the bispectrum/amplitude + closure phase images in quality as the number of closure amplitudes increases, even if some of those closure quantities contain zero baselines from co-located sites.

\subsection{Results: VLBA and ALMA  Images}
\label{sec::ALMA}

To test its performance on real observations, we applied our closure-only imaging algorithms on millimeter-wavelength interferometric datasets from the VLBA and ALMA. In both cases, the number of visibilities and closure quantities greatly exceed the number produced by the sparse EHT 2017 array, so we used NFFTs to speed up the imaging procedure. 
\begin{figure*}[hbtp]
\centering
\includegraphics*[width=0.33\textwidth]{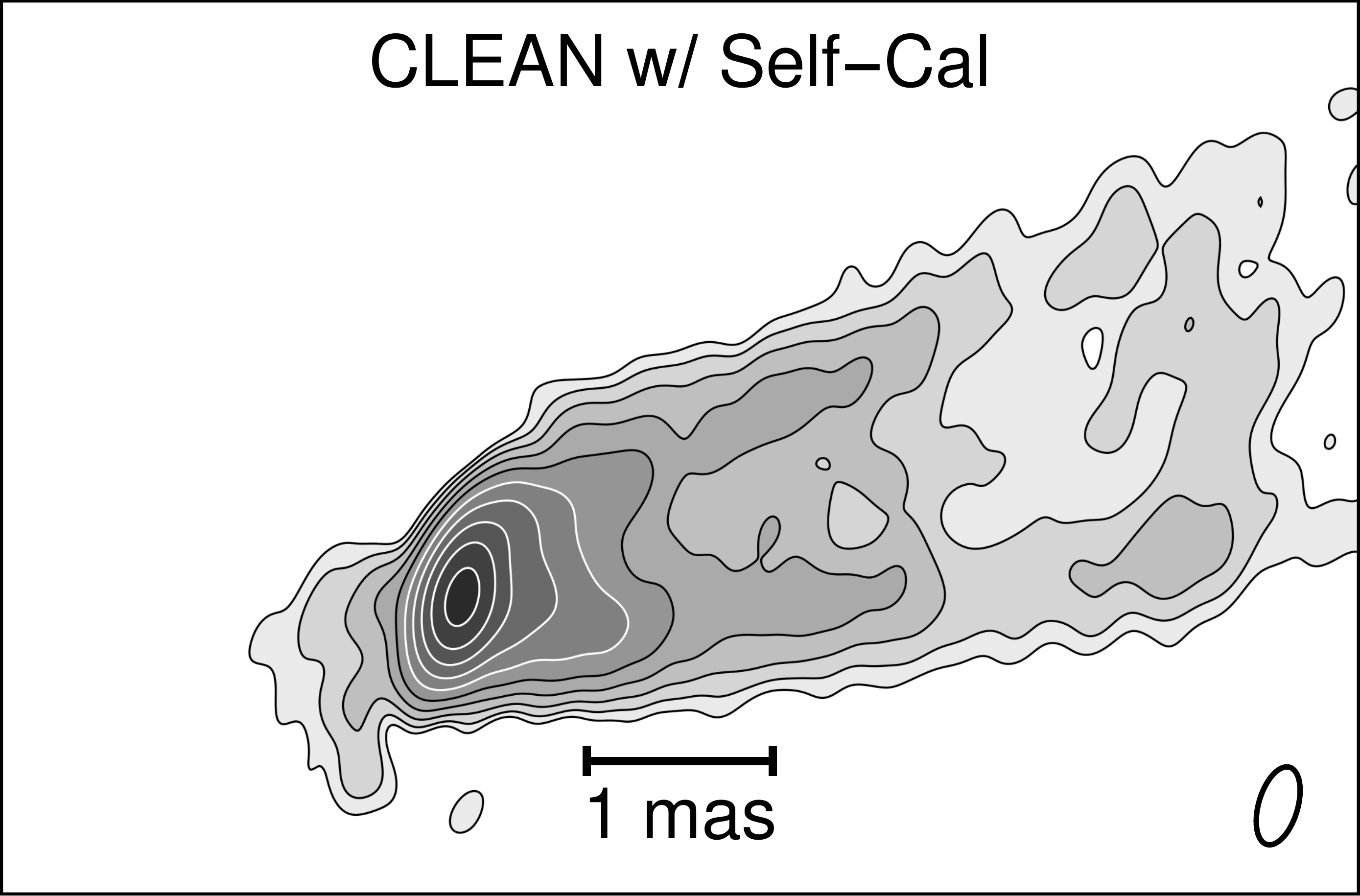}
\includegraphics*[width=0.33\textwidth]{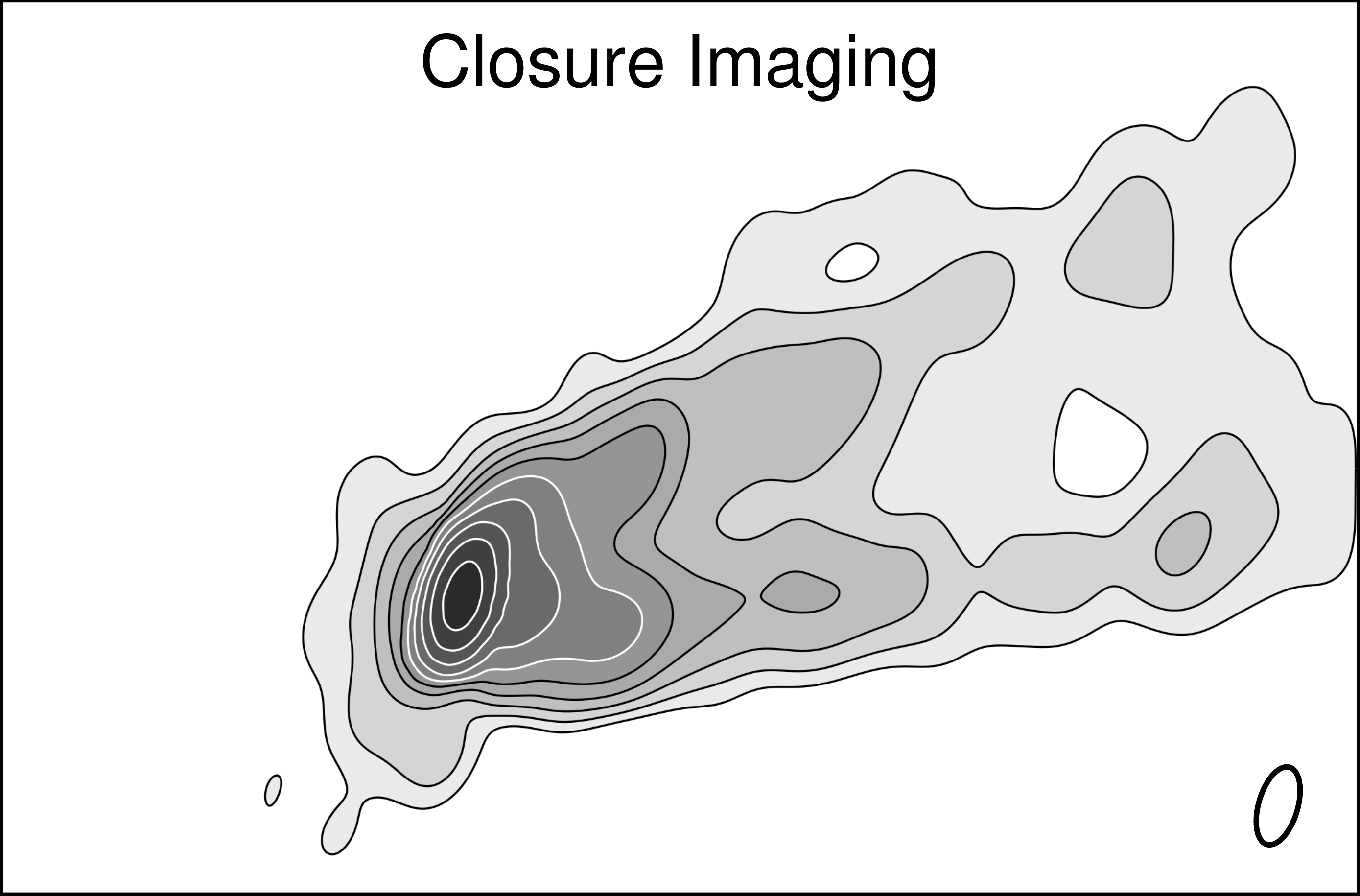}
\includegraphics*[width=0.33\textwidth]{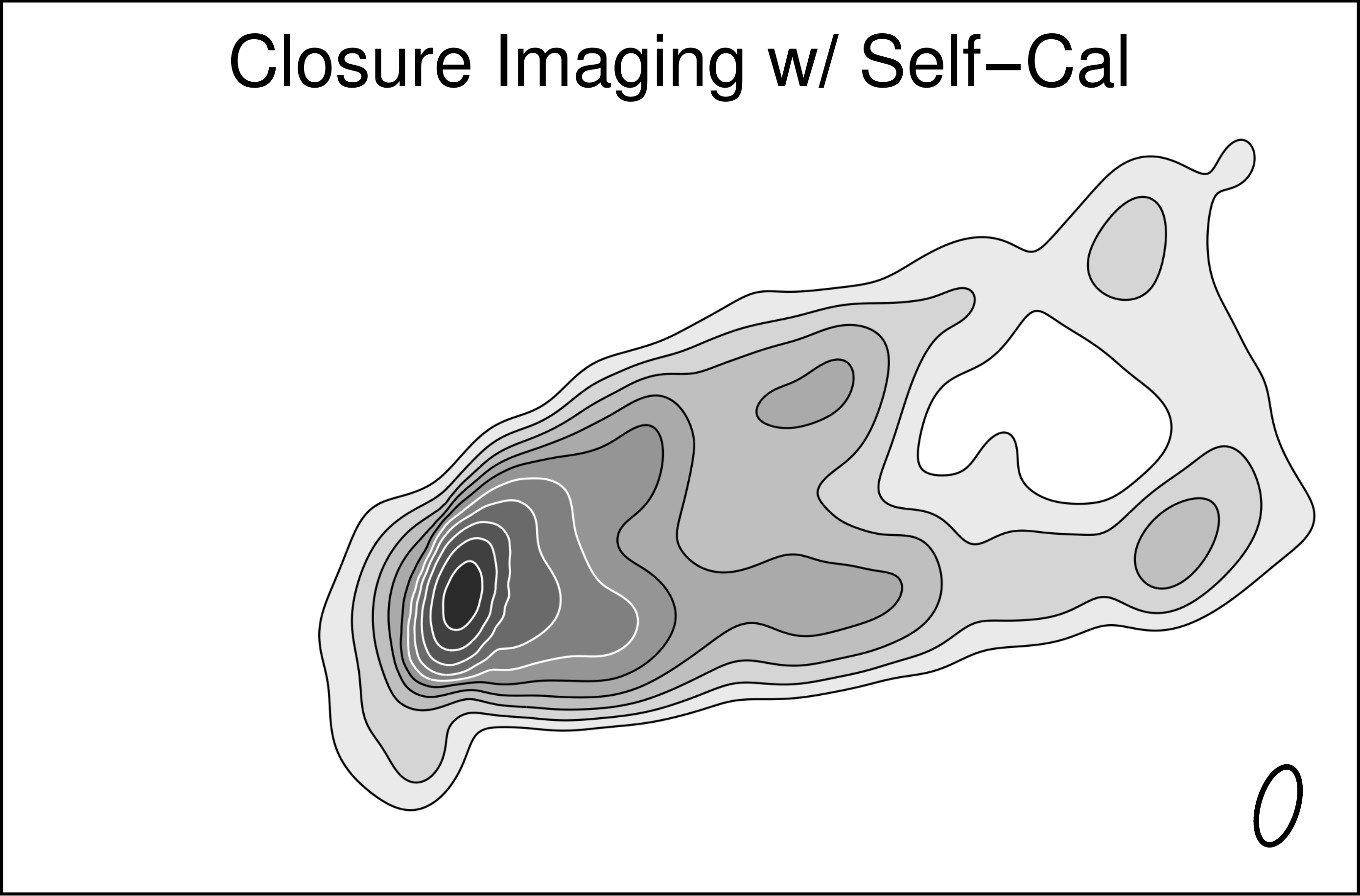}
\caption
{ 
Application of closure-only imaging to a VLBA observation of M87 at 7mm wavelength observed on May 9, 2007 \citep[for details, see][]{Walker_2016,Walker_2017}. (Left) CLEAN image made using iterative imaging and self-calibration. (Center) image reconstructed using closure-only imaging with a weak visibility amplitude constraint to aid initial convergence. (Right) image reconstructed using complex visibilities after self-calibrating to the closure-only image. To simplify the comparison between these approaches, each image has been convolved with the same CLEAN restoring beam and each image is rescaled to have the same total flux density as the CLEAN image. Contours in all panels are at equal levels, starting at $9.7~{\rm mJy}/{\rm mas}^2$ (${=}1~{\rm mJy}/{\rm beam}$) and increasing by factors of 2. 
}
\label{fig::M87}
\end{figure*}
Our first example is a VLBA observation of M87 at 7mm wavelength. In this case, and for other images with jets or narrow structure (see Fig.~\ref{fig::eht_dataset_3}), we have found that the major difficulty in closure-only imaging is convergence in the minimization of the objective function Equation~\eqref{eq::objfunc}. When we initialize to an uninformative image, the algorithm has difficulty converging to an image that has a reduced chi-squared near 1 in either the closure phases or the log closure amplitudes. 

To mitigate this problem while still preserving the benefits of only using closure quantities, we have found that initially including the visibility amplitudes in the minimization, even with a low weight and no calibration, can significantly aid the initial convergence. To avoid any bias from self-calibration in the data, we first applied a ``null'' calibration to the M87 data by assuming a single, constant SEFD (see Equation~\ref{eq::noise}) for all sites and all times. Next, we imaged the data using closure quantities and visibility amplitudes, which were down-weighted by a factor of 10 relative to the closure quantities. We then performed another two rounds toward convergence, initializing to the previous image convolved with a circular Gaussian matching the nominal array resolution, but with visibility amplitudes this time down-weighted by a factor of 100. Finally, we performed two additional rounds of imaging using only closure quantities. 

Figure~\ref{fig::M87} compares the reconstructed image to an image reconstructed using CLEAN and iterative self-calibration \citep{Walker_2016,Walker_2017}. We also derived a table of complex gains from a single iteration of self calibration to the final closure image to test how different our self-calibration solution would be from that obtained with CLEAN. The self-calibrated gains are significantly different than our initialized ``null'' calibration solution; after normalizing to the median gain (effectively fixing the total flux density), although 50\% of visibilities had residual gain corrections of less than 3\%, 10\% of the visibilities had residual gain corrections of more than 30\%. This result justifies post-hoc our choice to use visibility amplitudes in the initial minimization  steps. The majority of uncalibrated amplitudes have low error compared to the final self-calibrated set, so they are useful in aiding convergence; however, relying primarily on closure amplitudes ensures a final image that is less affected by the large gain errors present on some baselines. 

For comparison, we also applied our self-calibration solution to the data and then produced an image with the self-calibrated complex visibilities, minimizing Equation~\eqref{eq::objfunc} with a standard complex visibility $\chi^2$ term, Equation~\eqref{eq::rchisq}. The result is displayed in the third panel of Figure~\ref{fig::M87}. All three methods in Figure~\ref{fig::M87} give results that are broadly consistent, demonstrating the potential of closure imaging to obtain images that are comparable to those obtained by multiple rounds of finely-tuned CLEAN and self-calibration from an expert user. 

A general characteristic of closure imaging is its tendency to avoid high-frequency artifacts when highly converged; by removing spurious features from CLEAN images, closure methods could be useful in aiding in the physical interpretation of VLBI images. However, note that the CLEAN reconstruction in Figure~\ref{fig::M87} recovers more extended structure along the jet. This is likely because the CLEAN reconstructions were done using a multi-scale approach \citep{Wakker_88,Cornwell_08}; a similar multi-scale approach could likewise be used to improve closure imaging and is a key goal for improving the methods presented in this paper and applying them to further data sets.  

\begin{figure*}[htbp]
 \centering
 \includegraphics*[width=\linewidth]{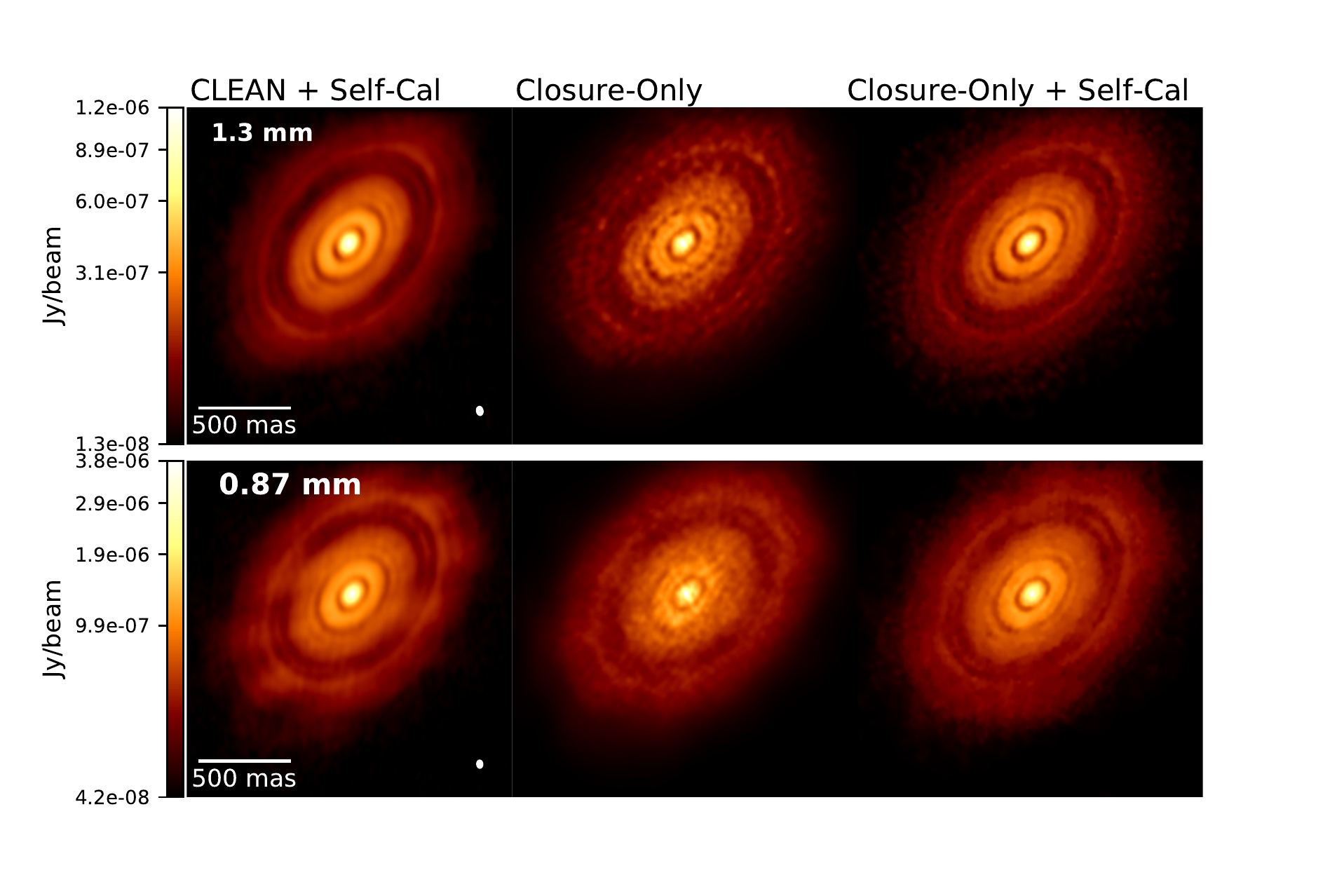}
 \caption
 {(Top) 1.3 mm band 6 ALMA image of the protoplanetary disk around HL Tau, comparing the CLEAN reconstruction from \citet{ALMA_HlTau} with our reconstructions. The leftmost panel shows the CLEAN image with a FOV of $1.8''$. The center panel shows an image of the same data produced by directly fitting to log closure amplitudes and closure phases, with downweighted visibility amplitudes used in the initial steps to aid convergence. Closure-only imaging produces an image that is consistent with the CLEAN result, despite not using any multi-scale imaging, but the overall resolution is lower. The rightmost panel shows an image produced from complex visibilities using a strong total variation regularizer after self-calibrating the data to the center closure-only image. After self-calibration, complex visibility imaging with total variation produces a sharp image with distinct disk gaps. 
 (Bottom) 0.87 mm band 7 ALMA images, produced using the same imaging parameters as the top 1.3 mm images. The 0.87 mm image obtained after closure-only imaging and one round of self-calibration eliminates prominent clean artifacts (dark spots) present in the original image. Our 0.87 mm image is similar to recently reprocessed images using CLEAN and a modified self-calibration loop \citep{Akiyama_2016}.
 }
\label{fig::hltau}
\end{figure*}

With 64 telescopes, ALMA has baseline coverage that much more densely fills the $u,v$ plane than the EHT and VLBA observations considered above. We imaged a 2014 ALMA observation of HL Tau taken both at 1.3 mm and 0.87 mm \citep{ALMA_HlTau} using our log closure amplitude and closure phase method described in \S\ref{sec::dterms} and \S\ref{sec::implementation}. We first averaged the data in five minute intervals. As in Figure~\ref{fig::M87}, we used downweighted visibility amplitudes in the initial steps of the minimization  to aid in convergence and removed them in the final runs of the imager. 

The results are displayed in Figure~\ref{fig::hltau}. Closure imaging is able to  replicate the overall structure of the published CLEAN images, including all of the gaps in the protoplanetary disk identified by the original reconstruction. Critically, our closure imaging algorithm does not yet include multi-scale imaging \citep{Wakker_88,Cornwell_08}, which was necessary to produce the ALMA CLEAN image. After producing an image from closure quantities, we self-calibrated the data to the resulting image (center panel of Figure~\ref{fig::hltau}) and imaged the data again directly using the resulting complex visibilities (minimizing Equation~\eqref{eq::objfunc} with Equation~\eqref{eq::rchisq}). The resulting image has a higher resolution than the closure-only image alone, with sharper and more distinct gaps apparent in the disk (right panel of Figure~\ref{fig::hltau}). Furthermore, particularly in the 0.87 mm image, the final reconstruction lacks the prominent periodic dark spots present in the CLEAN image that are likely caused by prominent dirty beam sidelobes, which were also ameliorated in recently reprocessed CLEAN + self-calibration images by \citet{Akiyama_2016}.

\section{Discussion}
\label{sec::Discussion}

The results in \S\ref{sec::EHT} and \S\ref{sec::ALMA} demonstrate that closure amplitudes and phases can be directly used in interferometric imaging to produce images that are insensitive to phase and amplitude calibration errors. Traditional self-calibration and imaging loops require many iterations of CLEAN imaging and fitting complex gains to visibilities \citep[e.g.,][]{Wilkinson77,Readhead78,Readhead80,Schwab_1980, Cornwell_1981,Pearson_1984,Walker_1995,Cornwell_1999}. These loops contain many tunable parameters including the choice of initial source model, the strategy for independent or concurrent calibration of amplitude  and phase gains, the CLEAN convergence criterion, the choice of taper and weighting for the CLEAN visibilities,  and the scales and regions to clean in each CLEAN iteration.

Our closure imaging method does not remove all tunable parameters from the model, but imaging with closure quantities alone necessarily produces results that are less biased by calibration assumptions. Images from closure imaging can stand on their own as minimal assumption estimates of the source structure; alternatively, results from closure-only imaging may be used as a well-motivated self-calibration model or initial source image for other imaging pipelines using calibrated data. On the ALMA and VLBA datasets in \S\ref{sec::ALMA}, we found that just one round of self-calibration to an image produced with closure quantities can be used to produce smooth high-resolution images from the resulting complex visibilities that match the best iterative, multi-scale CLEAN + self-calibration results.

The most significant challenge that we have encountered in closure-only imaging is difficulty in the early convergence and a tendency to quickly get stuck in wildly incorrect local minima. Counterintuitively, this tendency seems to be more of a problem in datasets with more interferometric baselines. This limitation may arise because the energy landscape represented by the closure amplitude terms (Equations~\ref{eq::chisqcamp} and \ref{eq::chisqgradlcamp}) becomes increasingly complicated with more correlated closure data. When using simulated data from the sparse EHT array (\S\ref{sec::EHT}), using closure quantities alone with a reasonable Gaussian prior and several imaging iterations is enough to guide the algorithm to converge on  a reasonable image.  However, imaging the real datasets from ALMA and the VLBA in \S\ref{sec::ALMA} using closure quantities alone with an uninformative initial model was difficult. For these cases, we found that adding a weak data constraint using uncalibrated visibility amplitudes (Equation~\ref{eq::rchisqamp}) helped guide the minimizer to the region of a good local minimum. This constraint can be as low as 1-10\% of the closure amplitude $\chi^2$ term and still produce excellent results; in practice, the amplitude error bars can also be increased to represent an estimate of the systematic amplitude error and further downweight this term \citep{Akiyama_Closure}. As the imaging proceeds, we remove the amplitude constraint and ultimately allow the final image to be only guided by the closure amplitudes and phases. Given the robustness of the results in Figures~\ref{fig::M87} and \ref{fig::hltau} to different choices of regularizer and data weights in the presence of a weak amplitude constraint, we see significant promise for this method to eventually allow for unsupervised closure imaging that can blindly produce a calibration-free image from decent initial data without user intervention.  

In the \texttt{eht-imaging} library, we have developed a flexible framework where images can be easily produced  from the same data set using different data and regularizer terms.  \texttt{eht-imaging} can also be used to self-calibrate data, to generate synthetic data from images with realistic thermal error and calibration uncertainties, and for the general plotting, analysis, and comparison of interferometer data.  Within this framework, it is easy to experiment with different arbitrary combinations of data terms and implement new imaging methods, such as polarimetric imaging \citep{Chael_16}, imaging in the presence of refractive scattering \citep{stochastic_optics}, and producing continuous movies from multi-epoch observations \citep{johnson_dyn}. 

\section{Conclusions}
\label{sec::Conclusions}
We have presented a framework for interferometric imaging using regularized maximum likelihood with arbitrary data products and its implementation in the software library \texttt{eht-imaging}. This work builds on decades of past work in applying regularized maximum likelihood approaches to interferometric imaging, and is in particular inspired by the simultaneous minimization of multiple data terms pioneered in optical interferometric imaging (see e.g. \citet{Thiebaut_2013, Thiebaut_2017}). This work extends that framework by imaging data directly with closure amplitudes (or their logarithms) for the first time,  rather than relying on amplitude self-calibration. 

In \S\ref{sec::gradients} and the Appendix, we gave analytic expressions for the gradients of various data $\chi^2$ terms including those for closure phases, closure amplitudes, and log closure amplitudes. The most powerful feature of this framework is its ability to produce images using closure quantities directly, making it possible to produce images directly from uncalibrated data. 

Using our method of closure-only imaging, self-calibration in imaging can be bypassed entirely, producing an image that will contain minimal calibration assumptions and will not depend on the choice of initial self-calibration model or other assumptions made in the self-calibration loop. In \S\ref{sec::EHT}, we showed that this strategy performs well on simulated EHT data of \sgra\ and M87. Images produced using only closure quantities have consistent fidelity at all levels of amplitude gain or miscalibration. Furthermore, when redundant sites are included in the array, the overall fidelity of the results approaches that of images made with perfectly calibrated data using conventional algorithms. 

In \S\ref{sec::ALMA}, we showed that closure imaging can also produce high quality images for VLBA and ALMA datasets at millimeter wavelengths, giving results that are of comparable quality to expert reconstructions with multi-scale CLEAN and self-calibration. Results from closure-only imaging can also be used to self-calibrate data and initialize additional imaging. We found that for the ALMA datasets considered, just one round of self-calibration and complex visibility imaging after closure-only imaging produces further refined results with fewer suspicious features that may be attributable to artifacts from CLEAN.  

Techniques involving calibration-insensitive closure quantities like those presented in this paper can help push interferometric imaging to more and more challenging regimes, including higher frequencies. While applicable to all interferometric astronomical data, our techniques are especially valuable at millimeter and sub-millimeter wavelengths, where calibration uncertainties are a large and variable component of the error budget. In \S\ref{sec::ALMA}, we have found that including a soft constraint from uncalibrated visibility amplitudes can dramatically aid in the convergence of closure imaging. Adding more data terms, this method can be easily generalized for polarimetric imaging \citep{Chael_16,akiyama_pol}, spectral index maps, simultaneous multi-band images, scattering mitigation \citep{stochastic_optics}, and dynamical movies of multi-epoch data \citep{starwarps,johnson_dyn}. 

\acknowledgements{We extend particular thanks to the anonymous referee, whose suggestion of the NFFT library and unusually thorough and thoughtful comments substantially improved this work. We thank Craig Walker and Michael Jan{\ss}en for their helpful conversations, comments and suggestions. We thank Roman Gold, Monica Mo{\'s}cibrodzka, and C.K. Chan for providing model images used in \S\ref{sec::EHT} and Svetlana Jorstad and Alan Marscher for the 43 GHz 3C273 image used in the same section. We additionally thank Craig Walker for providing the 7mm VLBA data used in \S\ref{sec::ALMA}. We thank the National Science Foundation (AST-1440254, AST-1614868, AST-1312651) and the Gordon and Betty Moore Foundation (GBMF-5278) for financial support of this work. This work was supported in part by the Black Hole Initiative at Harvard University, which is supported by a grant from the John Templeton Foundation. K.A. is financially supported by the Jansky Fellowship of the National Radio Astronomy Observatory. 
\\ \\
This study makes use of 43 GHz VLBA data from the VLBA-BU Blazar Monitoring Program (VLBA-BU-BLAZAR; \url{http://www.bu.edu/blazars/VLBAproject.html}), funded by NASA through the Fermi Guest Investigator Program. The VLBA is operated by the Long Baseline Observatory. The Long Baseline Observatory is a facility of the National Science Foundation operated under cooperative agreement by Associated Universities, Inc. This paper makes use of the following ALMA data: ADS/JAO.ALMA\#2011.0.00015.SV. ALMA is a partnership of ESO (representing its member states), NSF (USA) and NINS (Japan), together with NRC (Canada) and NSC and ASIAA (Taiwan), and KASI (Republic of Korea), in cooperation with the Republic of Chile. The Joint ALMA Observatory is operated by ESO, AUI/NRAO and NAOJ. The National Radio Astronomy Observatory is a facility of the National Science Foundation operated under cooperative agreement by Associated  Universities, Inc.}
\clearpage
\appendix
We present here the expressions for the gradients of the various data terms presented in \S\ref{sec::dterms} that we use in our imaging software. The equations below assume a DTFT matrix $A_{ij}$ (see Equation~\ref{eq::DTFTmatrix}); the conjugate transpose matrix $A^\dagger_{ij}$ gives the adjoint DTFT matrix. (note that since the visibility data is sparsely sampled, $A^\dagger A \neq 1$.)

The gradient of the complex visibility $\chi^2$ term (Equation~\ref{eq::rchisq}), already presented in the main text as Equation~\eqref{eq::chisqgradvis}, is
\begin{equation}
\label{eq::chisqgradvis2}
 \frac{\partial}{\partial I_i}\chi^2_\text{vis} = -\frac{1}{N}\sum_{j} \operatorname{Re}\left[A^\dagger_{ij}\left(\frac{V_j-V'_j}{\sigma^2_j}\right)\right].
\end{equation}

The gradient of the visibility amplitude $\chi^2$ (Equation~\ref{eq::rchisqamp}) is
\begin{equation}
 \frac{\partial}{\partial I_i}\chi^2_\text{amp} = -\frac{2}{N}\sum_{j} \operatorname{Re}\left[A^\dagger_{ij}\frac{V'_j}{|V'_j|}\left(\frac{\left(|V_j|-|V'_j|\right)}{\sigma^2_j}\right)\right].
\end{equation}

For the bispectrum $\chi^2$ (Equation~\ref{eq::chisqb}),  the gradient is
\begin{align}
 \frac{\partial}{\partial I_i}\chi^2_\text{bispec} = -\frac{1}{N_B}\sum_{j} \operatorname{Re} &\vphantom{+} \left[\left(\frac{A^\dagger_{1ij}}{V'^*_{1j}}+\frac{A^\dagger_{2ij}}{V'^*_{2j}}+\frac{A^\dagger_{3ij}}{V'^*_{3j}}\right)\left(\frac{(V_{Bj}-V'_{Bj})V'_{Bj}}{\sigma^2_{Bj}}\right)\right].
\end{align}

The closure phase $\chi^2$ (Equation~\ref{eq::chisqcphase}) has a gradient
\begin{align}
\frac{\partial}{\partial I_i}\chi^2_\text{cl phase} = -\frac{2}{N_{\psi}}\sum_{j} \operatorname{Im}&\vphantom{+}\left[\left(\frac{A^\dagger_{1ij}}{V'^*_{1j}}+\frac{A^\dagger_{2ij}}{V'^*_{2j}}+\frac{A^\dagger_{3ij}}{V'^*_{3j}}\right)\left(\frac{\sin(\psi_j-\psi'_j)}{\sigma^2_{\psi j}}\right)\right].
\end{align}

And finally, the gradients of the closure amplitude $\chi^2$ term (Equation~\ref{eq::chisqcamp}) is
\begin{align}
 \frac{\partial}{\partial I_i}\chi^2_\text{cl amp} = -\frac{2}{N_{C}}\sum_{j} \operatorname{Re} &\vphantom{+}\left[\left(\frac{A^\dagger_{1ij}}{V'^*_{1j}}+\frac{A^\dagger_{2ij}}{V'^*_{2j}} - \frac{A^\dagger_{3ij}}{V'^*_{3j}} - \frac{A^\dagger_{4ij}}{V'^*_{4j}}\right)\left(\frac{\left(|V'_{Cj}|-|V'_{Cj}|\right)|V'_{Cj}|}{\sigma^2_{Cj}}\right)\right],
\label{eq::chisqgradcamp}
\end{align}

and for log closure amplitudes (Equation~\ref{eq::chisqlcamp}) it is
\begin{align}
\label{eq::chisqgradlcamp}
 \frac{\partial}{\partial I_i}\chi^2_\text{log cl amp} = -\frac{2}{N_{C}}\sum_{j} \operatorname{Re} &\vphantom{+}\left[\left(\frac{A^\dagger_{1ij}}{V'^*_{1j}}+\frac{A^\dagger_{2ij}}{V'^*_{2j}} - \frac{A^\dagger_{3ij}}{V'^*_{3j}} - \frac{A^\dagger_{4ij}}{V'^*_{4j}}\right)\frac{|V'_{Cj}|^2}{\sigma^2_{Cj}}\log\left(\frac{\left|V_{Cj}\right|}{\left|V'_{Cj}\right|}\right)\right].
\end{align}

\newpage
\bibliography{ClosureOnly.bib}

\end{document}